\documentclass[prc,onecolumn,letterpaper,10pt,twoside,tightenlines,nofootinbib,showpacs,notitlepage]{revtex4-1}
\usepackage{amsmath,amsfonts,amssymb,latexsym}
\usepackage{graphicx}
\usepackage[sort&compress]{natbib}
\usepackage{grffile,epstopdf}
\usepackage{subfigure}
\usepackage{hhline}
\newcommand{\m}{\cdot}

\newcommand{\n}{\nonumber \\}

\newcommand{\llkl}{\left\langle}
\newcommand{\rrkl}{\right\rangle}

\newcommand{\kl}{\left(}
\newcommand{\kr}{\right)}

\newcommand{\sqs}{\sqrt{s}}
	
\makeindex

\begin{document}

\title{Electric Conductivity of a hot hadron gas from a kinetic approach}
\author{Moritz Greif}
\email{greif@th.physik.uni-frankfurt.de}
\affiliation{Institut f\"ur Theoretische Physik, Johann Wolfgang Goethe-Universit\"at,
Max-von-Laue-Str.\ 1, D-60438 Frankfurt am Main, Germany}
\author{Carsten Greiner}
\affiliation{Institut f\"ur Theoretische Physik, Johann Wolfgang Goethe-Universit\"at,
Max-von-Laue-Str.\ 1, D-60438 Frankfurt am Main, Germany}
\author{Gabriel S. Denicol}
\affiliation{Physics Department, Brookhaven National Lab, Building 510A, Upton, New York
11973, USA}
\date{\today }

\begin{abstract}
We calculate the electric conductivity of a gas of relativistic particles with isotropic cross sections using the
Boltzmann equation as the starting point. Our analyses is restricted to elastic collisions. We show the perfect agreement
with previously published numerical results for a massless quark-gluon
plasma, and give results for the electric conductivity of an interacting
hadron gas, employing realistic resonance cross sections. These results for
the electric conductivity of a hot hadron gas, as created in
(ultra-)relativistic heavy-ion collisions, are of rich phenomenological as
well as theoretical interest and can be compared to, e.g., lattice quantum field theory 
calculations.
\end{abstract}

\maketitle



\section{Introduction}

\label{sec:Intro}
In ultrarelativistic heavy ion collisions the core of the fireball can reach temperatures high enough 
to temporarily produce a new phase of nuclear matter, the quark-gluon plasma (QGP), in which quarks and gluons are the relevant degrees of freedom \cite{Arsene:2004fa,Adcox:2004mh,Back:2004je,Adams:2005dq}. After a few fm/c (in the center of momentum of the collision), the
nuclear matter produced cools down and undergoes a phase transition into a hadronic phase \cite{Kolb:2000sd}. The
hadron gas produced at the late stages of the collision is still hot, with temperatures $\lesssim 160~\mathrm{MeV}$, and
hadrons can still collide multiple times before they stream freely into the
detector. 

The theoretical understanding of the experimentally measured data
is essential for gaining knowledge of our nature at extreme scales. Among the
most successful descriptive pictures of the different phases of nuclear
matter are hydrodynamic calculations \cite{GALE2013,Schenke2011,PhysRevC.85.054902,Kolb:2003dz,Teaney:2001av,DelZanna:2013eua,Karpenko:2013wva,Holopainen:2010gz}, and
solutions of the Boltzmann equation (BE) \cite{Xu2005,Bouras2010a,Bouras2012,Bouras2009,Fochler2010,Fochler2011,Uphoff2011a,Wesp2011,Reining2012,Uphoff2012,Fochler2013,Greif2013,Senzel2013,Uphoff2013}. Hydrodynamical calculations describe the QGP \cite{Gyulassy2005} and the hadron gas (HG) as a droplet of viscous fluid, and need as an input macroscopic properties of the matter, as the equation of
state (EOS) and transport coefficients, like the shear and bulk viscosity.
The existence of a finite shear viscosity in the QGP is necessary to
explain, e.g., experimental data of the elliptic flow coefficient $v_2$ \cite{Schenke2011a}. The Boltzmann equation, governing the time development of a
particle distribution function due to collisions, allows for a direct
computation of transport coefficients, and also the space-time development
of the QGP phase can be described numerically \cite{Xu2005,Fochler2010,Fochler2011,Uphoff2011a,Uphoff2012,Fochler2013,Senzel2013,Uphoff2013}. In some of these studies, the BE was solved numerically in a fixed box, employing
various cross section for a given set of particles. With such a setup,
the transport coefficients shear viscosity over entropy density $\eta/s$,
heat flow $\kappa$, as well as the static electric conductivity $\sigma_%
\text{el}$ could be computed directly, see, e.g., Ref.~ \cite{Wesp2011,Reining2012,Plumari2012,Greif2013,Greif2014,Puglisi2014a,Puglisi2014b}. The computation of the latter coefficient in a new, analytic way is the
aim of this paper. Using established analytic developments \cite%
{Denicol2012b,Denicol2012,Denicol2010}, we investigated how an equilibrated
relativistic gas of electrically charged particles, governed by the BE,
behaves upon the influence of a small, static, electric field that is turned on. Assuming that the total system is electrically neutral,
naturally an electric current will develop and eventually reach a static
value (in an infinitely large system or setting periodic boundary
conditions). The longitudinal static electric conductivity $\sigma_\text{el}$
relates the response of the electric current\footnote{%
More precise, the electrically charged particle diffusion current density} $%
\vec{j}$ to the externally applied static electric field $\vec{E}$, 
\begin{equation}
\vec{j}=\sigma_{\text{el}}\vec{E}.
\end{equation}
We can thereby compute $\sigma_\text{el}$ for a given set of (massive or massless) particle
species in the system and the given set of their mutual, elastic, collision
cross section. This is basically an extension to the well-known Drude formula for the electric conductivity (see Sec.~\ref{sec:DrudeLimit}) for a hadron resonance gas.

The electric conductivity can be related to the soft dilepton production
rate \cite{Moore2003} and the diffusion of magnetic fields in a medium \cite%
{Baym1997,Tuchin2013,Fernandez-Fraile2006}.

Many scientific groups have recently investigated this transport
coefficient, including the mentioned numerical solution of the BE \cite%
{Greif2014,Puglisi2014b}, off-shell transport models \cite%
{Cassing2013,Steinert2013}, holography \cite{Finazzo2014}, lattice gauge
theory \cite{Aarts:2014JHEP,Aarts2007,Brandt2013,Amato2013a,Gupta2004,Buividovich2010,Burnier2012,Ding2011,Kaczmarek:2013dya}%
, Dyson-Schwinger calculations \cite{Qin2013}, a dynamical quasiparticle model \cite{Marty:2013ita,Berrehrah:2015vhe} and chiral perturbation theory 
\cite{Fernandez-Fraile2006}. All (but Ref.~\cite{Finazzo2014}) of these
calculations aim at the value of $\sigma_\text{el}$ in the QGP phase, some
do extend below the transition temperature towards the HG. In general, the
results differ over several orders of magnitude, and comparisons among
different approaches are often intriguing. 
In the HG there has been so far no analytic computation of the electric conductivity
from pure kinetic theory, something we will provide in this work. We
investigate the influence of masses, average total cross sections, and
different species. We finally state the temperature dependent electric
conductivity of a hadron gas with well justified approximations. Indeed, the
framework can give a very precise answer from kinetic theory for any (charge
neutral) elastic particle system, and is not restricted to the results
considered in this paper.

This work is organized as follows. In Sec.~\ref{basic_definitions} we give
basic definitions regarding the relativistic formulation for the fluid
dynamical quantities. In Sec.~\ref{sec:formalsism_ec} we derive the
algorithm for the computation of the conductivity from linear response, and
continue in Sec.~\ref{sec:Results} with our results. First, we reproduce
previously published numerical results and show the convergence of the
method in Sec.~\ref{sec:ResultsA}, then we show the influence of masses
systematically in Sec.\ref{sec:ResultsB}, followed by the results for a
realistic Pion gas in Sec.~\ref{sec:ResultsC}, a Pion-Nucleon-Kaon gas with
fixed cross sections (Sec.~\ref{sec:ResultsD}) and realistic cross sections
(Sec.~\ref{sec:ResultsE}). We give a conclusion and outlook in Sec.~\ref%
{sec:conclusion}.

Our units are $\hbar =c=k=1$; the space-time metric is given by $g^{\mu \nu
}=\text{diag}(1,-1,-1,-1)$. Greek indices run from $0$ to $3$. 

\section{Basic definitions}

\label{basic_definitions}

We consider a dilute gas consisting of $N_{\text{species}}$ particle
species, with the $i$-th particle species having electric charge $q_{i}$ and
degeneracy $g_{i}$. This system is in the presence of an external
electromagnetic field, given by an electromagnetic field strength $F^{\mu \nu
}$, and its net-electric charge density is assumed to be approximately zero at all
space-time points. The state of the system is characterized by the single
particle distribution function of each particle species, $f_{i}(x,p)$. The
time evolution equation satisfied by $f_{i}(x,p)$ is the Boltzmann equation,
which is an integro-differential equation with the following general
structure 
\begin{equation*}
k^{\mu }\frac{\partial }{\partial x^{\mu }}f_{\mathbf{k}}^{i}+k_{\nu
}q_{i}F^{\mu \nu }\frac{\partial }{\partial k^{\mu }}f_{\mathbf{k}%
}^{i}=\sum\limits_{j=1}^{N_{\text{species}}}C_{ij}(x^{\mu },k^{\mu }),
\end{equation*}%
where $C_{ij}$ is the collision term, that will be specified later in this
work. Since our goal is to calculate the electric conductivity of this
system, we shall consider the case of a homogeneous, but time-dependent
electric field.

The energy-momentum tensor and net electric charge four-current are expressed as the
following momentum integrals of the single-particle distribution function 
\begin{equation*}
T^{\mu \nu }=\sum\limits_{i=1}^{N_{\text{species}}}\left\langle k^{\mu
}k^{\nu }\right\rangle _{i},\quad N_{q}^{\mu }=\sum\limits_{i=1}^{N_{\text{%
species}}}q_{i}\left\langle k^{\mu }\right\rangle _{i}
\end{equation*}%
where we employ the following notation 
\begin{equation*}
\left\langle \ldots \right\rangle _{i}\equiv g_{i}\int \frac{d^{3}k}{(2\pi
)^{3}k^{0}}(\ldots )f_{\mathbf{k}}^{i}.
\end{equation*}%
These currents are associated to conserved quantities and satisfy the
continuity equations, $\partial _{\mu }T^{\mu \nu }=0$ and $\partial _{\mu
}N_{q}^{\mu }=0$.

It is convenient to decompose $T^{\mu \nu }$ and $N_{q}^{\mu }$ in terms of
the fluid's collective velocity field, $u^{\mu }$. Without loss of
generality, these currents are re-expressed as 
\begin{eqnarray}
T^{\mu \nu } &=&\epsilon u^{\mu }u^{\nu }-\Delta ^{\mu \nu }\left( P_{0}+\Pi
\right) +\pi ^{\mu \nu }, \\
N_{q}^{\mu }(x) &=&n_{q}u^{\mu }+V_{q}^{\mu }.  \label{Nmuformel}
\end{eqnarray}%
Above, we introduced the energy density $\epsilon $, the thermodynamic
pressure $P_{0}$, the bulk viscous pressure $\Pi $, the shear stress tensor $%
\pi ^{\mu \nu }$, the net electric charge density $n_{q}$, and the net electric charge
diffusion current $V_{q}^{\mu }$. We also defined the spatial projector $%
\Delta ^{\mu \nu }=g^{\mu \nu }-u^{\mu }u^{\nu }$ and employed Landau's
definition of the fluid velocity as an eigenvector of $T^{\mu \nu }$ with
eigenvalue $\epsilon $, that is, $T^{\mu \nu }u_{\nu }=\epsilon u^{\mu }$.
In this scheme, each new variable introduced is expressed by a given
contraction/projection of the currents with $u^{\mu }$ and $\Delta ^{\mu \nu
}$, 
\begin{eqnarray}
\epsilon &=&u_{\mu }u_{\nu }T^{\mu \nu }\text{, }P_{0}+\Pi =-\frac{1}{3}%
\Delta _{\mu \nu }T^{\mu \nu }\text{, } \\
\pi ^{\mu \nu } &=&\Delta _{\alpha \beta }^{\mu \nu }T^{\alpha \beta },\text{
}n_{q}=u_{\mu }N_{q}^{\mu }\text{, }V_{q}^{\mu }=N_{q}^{\left\langle \mu
\right\rangle }.
\end{eqnarray}%
For convenience, we adopt the notation $A^{\left\langle \mu \right\rangle
}\equiv \Delta _{\nu }^{\mu }A^{\nu }$ and $A^{\left\langle \mu \nu
\right\rangle }\equiv \Delta _{\alpha \beta }^{\mu \nu }A^{\alpha \beta }$.
The latter definition used the double, traceless, symmetric projection
operator $\Delta _{\alpha \beta }^{\mu \nu }=(\Delta _{\alpha }^{\mu }\Delta
_{\beta }^{\nu }+\Delta _{\alpha }^{\nu }\Delta _{\beta }^{\mu })/2-\Delta
^{\mu \nu }\Delta _{\alpha \beta }/3$. Since our goal will be to compute
the electric conductivity coefficient of a gas, most of the dissipative
currents introduced above will play no role in our calculation.
Nevertheless, we introduced them above for the sake of completeness.

We can define a temperature and chemical potential for this system using the
traditional matching conditions,
\begin{equation}
\epsilon =\epsilon ^{\mathrm{eq}}(T,\mu _{q})\text{, }n_{q}=n_{q}^{\mathrm{eq%
}}(T,\mu _{q}).
\end{equation}%
where $\epsilon ^{\mathrm{eq}}$ and $n_{q}^{\mathrm{eq}}$ are the energy
density and net electric charge density of a system in thermodynamic equilibrium with
temperature $T$ and chemical potential $\mu _{q}$. The values of temperature
and chemical potential must be inverted from the above equations. With these
definitions, we can introduce the local equilibrium distribution function, 
\begin{equation*}
f_{0,\mathbf{k}}^{i}=g_{i}\exp \left( -u_{\mu }k^{\mu }/T+q_{i}\mu
_{q}/T\right) ,
\end{equation*}%
and the deviation from equilibrium $\delta f_{\mathbf{k}}^{i}=f_{\mathbf{k}%
}^{i}-f_{0,\mathbf{k}}^{i}$, where, $\mu _{i}=q_{i}\mu _{q}$ is the chemical
potential of the $i$--th species. Momentum integrals over these distribution
functions will be expressed using the following notation 
\begin{equation*}
\left\langle \ldots \right\rangle _{i,0}\equiv g_{i}\int \frac{d^{3}k}{(2\pi
)^{3}k^{0}}(\ldots )f_{0,\mathbf{k}}^{i},\quad \left\langle \ldots
\right\rangle _{i,\delta }\equiv g_{i}\int \frac{d^{3}k}{(2\pi )^{3}k^{0}}%
(\ldots )\delta f_{\mathbf{k}}^{i}.
\end{equation*}
The electric net charge diffusion current then is 
\begin{equation*}
j^{\mu }=N_{q}^{\left\langle \mu \right\rangle }=\sum\limits_{i=1}^{N_{\text{%
species}}}q_{i}\left\langle k^{\mu }\right\rangle _{i,\delta }.
\end{equation*}


\section{Linear response to the electric field}

\label{sec:formalsism_ec} The scenario we want to consider here is that of a
thermal 'brick' of matter, in which the temperature $T\equiv \beta _{0}^{-1}$
and chemical potential $\mu _{q}\equiv \alpha _{0}^{q}/\beta _{0}$ do not
vary in space nor time. We generalize the methods proposed in \cite{Denicol:2011fa,Denicol:2011ef,Denicol:2013nua} to calculate retarded
Green's function associated to the response of a multi-component system to
an external electric field. We present the general calculation first, using
the full linearized collision term, and show afterwards that the formalism
reduces to the well-known Drude formula in the relaxation time
approximation. In all remaining computations we use the full linearized collision term.

\subsection{General calculation with linearized collision term}
We consider a system initially in thermal equilibrium, with $f_{\mathbf{k}%
}^{i}=f_{0,\mathbf{k}}^{i}$ and $F^{\mu \nu }=0$. We then suddenly turn on a
small external electric field. No external magnetic fields are present and
we neglect the effect of any induced field. The distribution function
acquires an off-equilibrium part, $f_{\mathbf{k}}^{i}=f_{0,\mathbf{k}%
}^{i}+\delta f_{\mathbf{k}}^{i}$, and the field strength tensor becomes 
\begin{equation}
F^{\mu \nu }\rightarrow \delta F^{\mu \nu }=E^{\mu }u^{\nu }-E^{\nu }u^{\mu }
\end{equation}%
where $E^{\mu }=u_{\nu }F^{\mu \nu }$ is the electric field. We write down
the linearized BE (similar to \cite{Denicol:2013nua}), neglecting any term
that is second order in $\delta f$, $\delta F^{\mu \nu }$, or their product, 
\begin{equation}
k^{\mu }\frac{\partial }{\partial x^{\mu }}f_{0,\mathbf{k}}^{i}+k^{\mu }%
\frac{\partial }{\partial x^{\mu }}\delta f_{\mathbf{k}}^{i}+k_{\nu
}q_{i}\delta F^{\mu \nu }\frac{\partial }{\partial k^{\mu }}f_{0,\mathbf{k}%
}^{i}=\sum\limits_{j=1}^{N_{\text{species}}}C_{ij}^{(l)}(x^{\mu },k^{\mu }),
\end{equation}%
with $C_{ij}^{(l)}(x^{\mu },k^{\mu })$ being the linearized collision term.
Without loss of generality, we carry out all computations in the local rest
frame of the fluid, $u^{\mu }=(1,0,0,0)$. Since $E^{\mu }$ is orthogonal to
the velocity, $u_{\mu }E^{\mu }=0$, we replace $k_{\nu }E^{\nu }\rightarrow
k_{\left\langle \nu \right\rangle }E^{\nu }$. Then we have 
\begin{equation}
k^{\mu }\frac{\partial }{\partial x^{\mu }}\delta f_{\mathbf{k}}^{i}+\frac{%
q_{i}}{T}f_{0,\mathbf{k}}^{i}k_{\left\langle \nu \right\rangle }E^{\nu
}=\sum\limits_{j=1}^{N_{\text{species}}}C_{ij}^{(l)}(x^{\mu },k^{\mu }).
\label{eq:BE_after_Efield}
\end{equation}
The linearized collision term can be written as an operator $\hat{C}$ acting
on $\delta f$, 
\begin{equation}
C_{ij}^{(l)}(x^{\mu },k^{\mu })\equiv \hat{C}\delta f_{\mathbf{k}}^{i}=\int 
\mathrm{d}K^{\prime }\mathrm{d}P\mathrm{d}P^{\prime }\gamma _{ij}W_{\mathbf{k%
}\mathbf{k}^{\prime }\rightarrow \mathbf{p}\mathbf{p}^{\prime }}^{ij}f_{0,%
\mathbf{k}}^{i}f_{0,\mathbf{k}^{\prime }}^{j}\left( \frac{\delta f_{\mathbf{p%
}}^{i}}{f_{0,\mathbf{p}}^{i}}+\frac{\delta f_{\mathbf{p}^{\prime }}^{i}}{%
f_{0,\mathbf{p}^{\prime }}^{j}}-\frac{\delta f_{\mathbf{k}}^{i}}{f_{0,%
\mathbf{k}}^{i}}-\frac{\delta f_{\mathbf{k}^{\prime }}^{j}}{f_{0,\mathbf{k}%
^{\prime }}^{j}}\right)   \label{eq:line_coll_term}
\end{equation}%
where we use the notation $dK\equiv \mathrm{d}^{3}k/\left[ (2\pi )^{3}k^{0}%
\right] $, $\gamma _{ij}=1-1/2\delta _{ij}$ and $W_{\mathbf{k}\mathbf{k}%
^{\prime }\rightarrow \mathbf{p}\mathbf{p}^{\prime }}^{ij}=s\sigma
_{ij}(s,\Theta )(2\pi )^{6}\delta ^{(4)}(k^{\mu }+k^{\prime \mu }-p^{\mu
}-p^{\prime \mu })$. Above, we only considered elastic 2-to-2 collisions. The
total cross section $\sigma _{\text{tot},ij}(s)$ is related to the
differential cross section $\sigma _{ij}(s,\Theta )$ in the following way, 
\begin{equation}
\sigma _{\text{tot},ij}(s)=\frac{2\pi }{\nu }\int \mathrm{d}\Theta \sin
\Theta \sigma _{ij}(s,\Theta ),\quad \cos \Theta =\frac{(k-k^{\prime
})(p-p^{\prime })}{(k-k^{\prime })^{2}},\quad s=(k+k^{\prime })^{2}.
\end{equation}

We take the Fourier transform of the Eq.~\eqref{eq:BE_after_Efield}, and
divide it by the energy $E_{\mathbf{k}}=\sqrt{\mathbf{k}^{2}+m^{2}}$,
leading to the following equation for the Fourier transform of the
nonequilibrium distribution function, $\delta \tilde{f}_{\mathbf{k}}^{i}$, 
\begin{gather}
-i\omega \delta \tilde{f}_{\mathbf{k}}^{i}+i\frac{\mathbf{k}}{E_{\mathbf{k}}}%
\cdot \mathbf{q}\delta \tilde{f}_{\mathbf{k}}^{i}-\sum\limits_{j=1}^{N_{%
\text{species}}}\frac{1}{E_{\mathbf{k}}}\hat{C}_{ij}\delta \tilde{f}_{%
\mathbf{k}}^{i}=-\frac{q_{i}}{TE_{\mathbf{k}}}f_{0,\mathbf{k}%
}k^{\left\langle \nu \right\rangle }\tilde{E}_{\nu }  \notag \\
\Rightarrow \delta \tilde{f}_{\mathbf{k}}^{i}=-\frac{1}{T}\frac{q_{i}}{%
-i\omega +i\frac{\mathbf{k}}{E_{\mathbf{k}}}\cdot \mathbf{q}-\sum_{j=1}^{N_{%
\text{species}}}\frac{1}{E_{\mathbf{k}}}\hat{C}_{ij}}f_{0,\mathbf{k}}^{i}%
\frac{k_{\left\langle \nu \right\rangle }}{E_{\mathbf{k}}}\tilde{E}^{\nu },
\label{eq:el_current}
\end{gather}%
where $\tilde{E}_{\nu }$ is the Fourier transform of $E_{\nu }$ and the last
equation is the formal solution for the distribution function in Fourier
space. Using the formal solution derived for $\delta \tilde{f}_{\mathbf{k}%
}^{i}$ in Eq.~\eqref{eq:el_current}, we can express the Fourier transform of
the net electric charge current in the following simple form 
\begin{equation}
\tilde{j}^{\mu }=-\sum_{i=1}^{N_{\text{species}}}\frac{q_{i}}{T}\int \mathrm{%
d}Kk^{\left\langle \mu \right\rangle }\frac{q_{i}}{-i\omega +i\frac{\mathbf{k%
}}{E_{\mathbf{k}}}\cdot \mathbf{q}-\sum_{j=1}^{N_{\text{species}}}\frac{1}{%
E_{\mathbf{k}}}\hat{C}_{ij}}f_{0,\mathbf{k}}^{i}\frac{k_{\left\langle \nu
\right\rangle }}{E_{\mathbf{k}}}\tilde{E}^{\nu }\equiv \tilde{G}_{R}^{\mu
\nu }(\omega ,\mathbf{q})\tilde{E}_{\nu },
\label{eq:current_definition_greens_function0}
\end{equation}
where we introduced the retarded Green's Function $\tilde{G}_{R}^{\mu \nu
}(\omega ,\mathbf{q})$.

In order to compute the static electric conductivity, it will be enough to
compute the retarded Greens function $\tilde{G}_{R}^{\mu \nu }(\omega ,%
\mathbf{q})$ at vanishing frequency and wavenumber, $\tilde{G}_{R}^{\mu \nu
}(0,\mathbf{0})$. For this purpose, we introduce a vector $B_{i}^{\alpha
}(Q,K_{i})$, which satisfies the following integro-differential equation 
\begin{equation}
\left[ -i\omega +i\frac{\mathbf{k}}{E_{\mathbf{k}}}\cdot \mathbf{q}%
-\sum_{j=1}^{N_{\text{species}}}\frac{1}{E_{\mathbf{k}}}\hat{C}_{ij}\right]
B_{i}^{\alpha }(Q,K_{i})=q_{i}f_{0,\mathbf{k}}^{i}\frac{k^{\left\langle
\alpha \right\rangle }}{E_{\mathbf{k}}}.
\label{eq:integro_diff_equation_for_B}
\end{equation}%
Once $B^{\alpha }$ is known, the solution for\ $\tilde{G}_{R}^{\mu \nu
}(\omega ,\mathbf{q})$ follows trivially as
\begin{equation}
\tilde{G}_{R}^{\mu \nu }(\omega ,\mathbf{q})=-\sum_{i=1}^{N_{\text{species}}}
\frac{q_{i}}{T}\int \mathrm{d}Kk^{\left\langle \mu \right\rangle
}B_{i}^{\left\langle \nu \right\rangle }(Q,K_{i}).
\label{eq:current_definition_greens_function}
\end{equation}%
Strictly speaking, $B^{\alpha }$ is a general function of $Q=(\omega ,%
\mathbf{k}),$ however, since we will need it only at vanishing $Q$, it is
sufficient to only consider its dependence on the 4-momentum $K$, that is, $%
B_{i}^{\alpha }(Q=0,K_{i})$. We know that $B_{i}^{\alpha }(K)$ is a 4-vector
orthogonal to $u^{\mu }$ and its tensor structure must be constructed from
combinations of $u^{\mu }$, $k^{\mu }$, and $g^{\mu \nu }$. Therefore, it
must be a tensor of the following form, $B_{i}^{\alpha }(K)\sim
k^{\left\langle \alpha \right\rangle }$, with the proportionality factors
being functions of the scalars $\mu _{q}$, $T$, and $E_{\mathbf{k}}$. It is
convenient to express it as an expansion in powers of the energy, 
\begin{equation}
B_{i}^{\alpha }(K)=f_{0,\mathbf{k}}^{i}k^{\left\langle \alpha \right\rangle
}\sum\limits_{n=0}^{\infty }a_{n}^{(i)}E_{\mathbf{k}}^{n},
\label{eq:expansion}
\end{equation}%
where $a_{n}^{(i)}$ are the expansion coefficients. Using the well-known
relation 
\begin{equation}
\int \mathrm{d}Kk^{\left\langle \mu \right\rangle }k^{\left\langle \nu
\right\rangle }E_{i,\mathbf{k}}^{n}f_{0,\mathbf{k}}^{i}=\frac{1}{3}\Delta
^{\mu \nu }\int \mathrm{d}KE_{\mathbf{k}}^{n}f_{0,\mathbf{k}}^{i}\Delta
_{\alpha \beta }k^{\alpha }k^{\beta },
\end{equation}%
together with Eqs.~\eqref{eq:current_definition_greens_function}~and 
\eqref{eq:expansion}, it is possible to express the
retarded Green's function in terms of the coefficients $a_{n}^{(i)}$,
\begin{equation}
\tilde{G}_{R}^{\mu \nu }(0,\mathbf{0})=-\Delta ^{\mu \nu }\sum_{i=1}^{N_{%
\text{species}}}\sum\limits_{n=0}^{\infty }\frac{q_{i}}{3T}a_{n}^{(i)}\int 
\mathrm{d}Kf_{0,\mathbf{k}}^{i}E_{\mathbf{k}}^{n}\Delta _{\alpha \beta
}k^{\alpha }k^{\beta }\equiv \Delta ^{\mu \nu }\tilde{G}_{R}.
\end{equation}%
Above, we defined the scalar retarded Green's function 
\begin{equation*}
\tilde{G}_{R}=-\sum_{i=1}^{N_{\text{species}}}\sum\limits_{n=0}^{\infty }%
\frac{q_{i}}{3T}a_{n}^{(i)}\int \mathrm{d}KE_{\mathbf{k}}^{n}(\Delta _{\mu
\nu }k^{\mu }k^{\nu })f_{0,\mathbf{k}}^{i},
\end{equation*}%
which can be used to express the linear relation between current and driving
electric field at $Q=0$ as 
\begin{equation*}
\tilde{j}^{\mu }=\tilde{G}_{R}\tilde{E}^{\mu }.
\end{equation*}%
The above relation allows us to identify the electric conductivity as $%
\sigma _{\text{el}} \equiv \tilde{G}_{R}$.

Naturally, the expansion \eqref{eq:expansion} must be truncated at some point
and we will discuss the convergence of our results to the order of the
truncation. We note that, even at the lowest possible order of truncation,
the resulting transport coefficients are expected to be accurate up to $10~\%
$, see, e.g., \cite{DeGroot,Denicol2012b}. Our next step is the
determination of the expansion coefficients $a_{n}^{(i)}$. Multiplying Eq.~%
\eqref{eq:integro_diff_equation_for_B} with $E_{\mathbf{k}%
}^{m}k^{\left\langle \beta \right\rangle }$ and integrating over momentum we
get an equation for $a_{n}^{(i)}$, 
\begin{equation*}
\sum\limits_{n=0}^{\infty }\int \mathrm{d}K_{i}E_{\mathbf{k}%
}^{m-1}k^{\left\langle \beta \right\rangle }\left[ -\sum_{j=1}^{N_{\text{%
species}}}\hat{C}_{ij}f_{0,\mathbf{k}}^{i}E_{\mathbf{k}}^{n}k^{\left\langle
\alpha \right\rangle }a_{n}^{(i)}\right] =q_{i}\int \mathrm{d}K_{i}E_{%
\mathbf{k}}^{m-1}k^{\left\langle \alpha \right\rangle }k^{\left\langle \beta
\right\rangle }f_{0,\mathbf{k}}^{i}.
\end{equation*}
Using straightforward manipulations of this equation and the above
definition of the collision term, Eq.~\eqref{eq:line_coll_term}, we can
rewrite it in the following form, 
\begin{equation*}
\sum\limits_{n=0}^{\infty }\sum_{j=1}^{N_{\text{species}}}\left[ \mathcal{A}%
_{mn}^{i}\delta ^{ij}+\mathcal{C}_{mn}^{ij}\right] a_{n}^{(j)}=b_{m}^{i},
\end{equation*}
where we defined 
\begin{align}
\mathcal{A}_{mn}^{i}& =\sum_{j=1}^{N_{\text{species}}}\int \mathrm{d}K_{i}%
\mathrm{d}K_{j}^{\prime }\mathrm{d}P_{i}\mathrm{d}P_{j}^{\prime }\gamma
_{ij}W_{\mathbf{k}\mathbf{k}^{\prime }\rightarrow \mathbf{p}\mathbf{p}%
^{\prime }}^{ij}f_{0,\mathbf{k}}^{i}f_{0,\mathbf{k}^{\prime }}^{j}E_{i,%
\mathbf{k}}^{m-1}k_{\left\langle \alpha \right\rangle }\left( E_{i,\mathbf{p}%
}^{n}p^{\left\langle \alpha \right\rangle }-E_{i,\mathbf{k}%
}^{n}k^{\left\langle \alpha \right\rangle }\right) ,  \notag \\
\mathcal{C}_{mn}^{ij}& =\int \mathrm{d}K_{i}\mathrm{d}K_{j}^{\prime }\mathrm{%
d}P_{i}\mathrm{d}P_{j}^{\prime }\gamma _{ij}W_{\mathbf{k}\mathbf{k}^{\prime
}\rightarrow \mathbf{p}\mathbf{p}^{\prime }}^{ij}f_{0,\mathbf{k}}^{i}f_{0,%
\mathbf{k}^{\prime }}^{j}E_{i,\mathbf{k}}^{m-1}k_{\left\langle \alpha
\right\rangle }\left( E_{j,\mathbf{p}^{\prime }}^{n}p^{\prime \left\langle
\alpha \right\rangle }-E_{i,\mathbf{k}^{\prime }}^{n}k^{\prime \left\langle
\alpha \right\rangle }\right) ,  \notag \\
b_{m}^{i}& =q_{i}\int \mathrm{d}KE_{\mathbf{k}}^{m-1}\left( -\Delta ^{\mu
\nu }k_{\mu }k_{\nu }\right) f_{0,\mathbf{k}}^{i}.  \label{eq:integrals}
\end{align}%
For later use we denote the above matrix in particle species space and
expansion space as 
\begin{equation}
\mathcal{N}_{mn}^{ij}\equiv \mathcal{A}_{mn}^{i}\delta ^{ij}+\mathcal{C}%
_{mn}^{ij}.
\label{eq:final_matrix}
\end{equation}%
Note that there is no sum over $i$ implied. The Landau matching condition
can also be expressed as 
\begin{equation*}
\Delta _{\nu }^{\lambda }u_{\mu }T^{\mu \nu }=\sum_{i=1}^{N_{\text{species}%
}}\int \mathrm{d}Ku_{\nu }k^{\nu }k^{\left\langle \mu \right\rangle }\delta
f_{\mathbf{k}}^{i}=-\frac{1}{3T}\sum_{i=1}^{N_{\text{species}%
}}\sum\limits_{n=0}^{\infty }a_{n}^{(i)}\int \mathrm{d}Kf_{0,\mathbf{k}%
}^{i}E_{\mathbf{k}}^{n+1}(\Delta ^{\alpha \beta }k_{\alpha }k_{\beta })%
\tilde{E}^{\mu }=0.
\end{equation*}
Since this should be true for any electric field and any of its components,
we obtain a constraint that must be satisfied by the coefficients $%
a_{n}^{(i)}$, 
\begin{align}
\sum_{i=1}^{N_{\text{species}}}\sum\limits_{n=0}^{\infty }a_{n}^{(i)}\left[
\int \mathrm{d}Kf_{0,\mathbf{k}}^{i}E_{\mathbf{k}}^{n+1}(\Delta ^{\alpha
\beta }k_{\alpha }k_{\beta })\right] & =0  \notag \\
\Rightarrow \sum_{i=1}^{N_{\text{species}}}\sum\limits_{n=0}^{\infty
}a_{n}^{(i)}d_{n}^{i}& =0\quad \text{with}\quad d_{n}^{i}\equiv \int \mathrm{%
d}Kf_{0,\mathbf{k}}^{i}E_{\mathbf{k}}^{n+1}(\Delta ^{\alpha \beta }k_{\alpha
}k_{\beta }).  \label{eq:additional_condition}
\end{align}

Solving the integrals in Eq.~\eqref{eq:integrals} for a given set of species
and cross sections allows us to obtain the unknown coefficients $a_{n}^{(i)}$
by inverting the matrix $\mathcal{A}_{mn}^{i}\delta ^{ij}+\mathcal{C}%
_{mn}^{ij}$ along with condition \eqref{eq:additional_condition}. In
practice, this amounts to removing one line and column from the matrix $%
\mathcal{N}_{mn}^{ij}$.

\subsection{Relaxation time limit}
\label{sec:DrudeLimit}
Nonrelativistically, the Drude formula for the electric conductivity $\sigma
_{\text{el,nr}}$ of a single charge carrying species (e.g. electrons) with
charge $q_e$, density $n_{e}$ and mass $m_{e}$ reads 
\begin{equation}
\sigma _{\text{el,nr}}=\frac{n_{e}q_e^{2}\tau }{m_{e}},
\label{Non-relativistic-Drude}
\end{equation}%
where $\tau $ is the mean time between collisions of the charge carriers
(e.g. electrons) with, e.g., atomic cores. The Boltzmann equation can be
solved analytically in the relaxation time approximation, which corresponds
to a simplistic model for the collision term, 
\begin{equation}
p^{\mu }\partial _{\mu }f_{q}+qF^{\alpha \beta }p_{\beta }\frac{\partial
f_{q}}{\partial p^{\alpha }}=-\frac{p^{\mu }u_{\mu }}{\tau }\left( f_{q}-f_{%
\text{eq},q}\right) .  \label{AndersonWittigModel}
\end{equation}%
It allows for a straightforward calculation of the charged particle
distribution $f_{q}$ after applying an external electric field. The
uncharged particle distribution remains thermal $f_{q=0}=f_{\text{eq},q=0}$
and is not affected by the electric field, 
\begin{equation}
\sigma _{\text{el}}=\frac{1}{3T}\sum_{i=1}^{N_{\text{species}%
}}q_{i}^{2}n_{i}\tau .  \label{EasyRelaxationFormula}
\end{equation}%
Here, $\tau $ is the mean time between collisions of particles, independent
of the particle type; for more details, see, e.g., Ref.~\cite{Greif2014}.
Using Eq.~\eqref{eq:current_definition_greens_function} with a relaxation
time collision operator we recover the relaxation time answer, Eq.~%
\eqref{EasyRelaxationFormula}, for the electric conductivity, 
\begin{align}
\tilde{j}^{\mu }& =\sum_{i=1}^{N_{\text{species}}}\frac{(q_{i})^{2}}{T}\int 
\mathrm{d}Kk^{\left\langle \mu \right\rangle }\frac{1}{-\sum_{j=1}^{N_{\text{%
species}}}\hat{C}_{ij}}f_{0,\mathbf{k}}^{i}k^{\left\langle \nu \right\rangle
}\tilde{E}_{\nu }  \notag \\
& =\sum_{i=1}^{N_{\text{species}}}\frac{(q_{i})^{2}}{T}\int \mathrm{d}%
Kk^{\left\langle \mu \right\rangle }\frac{\tau }{E_{\mathbf{k}}}f_{0,\mathbf{%
k}}^{i}k^{\left\langle \nu \right\rangle }\tilde{E}_{\nu }  \notag \\
& =\sum_{i=1}^{N_{\text{species}}}\frac{(q_{i})^{2}\tau }{3T}\left[ \int 
\mathrm{d}K\frac{1}{E_{\mathbf{k}}}(\Delta ^{\alpha \beta }k_{\alpha
}k_{\beta })f_{0,\mathbf{k}}^{i}\right] \tilde{E}^{\mu }  \notag \\
& =\sum_{i=1}^{N_{\text{species}}}\frac{(q_{i})^{2}\tau }{3T}n_{0,i}\tilde{E}%
^{\mu }.
\end{align}



\section{Results}
\label{sec:Results} 
Our main goal is to calculate the electric conductivity
of a hadron gas characterized by (measured) hadron-hadron cross sections
(e.g. Breit-Wigner peaked resonances). In practice we have to limit the
calculation to the dominant hadron species, such as pions, protons,
neutrons, kaons. To understand the results and to cross check our method, we
work systematically and include more species, masses and cross sections
step-by-step.
The use of simplified hadronic cross sections is common practise, e.g. in Ref.~\cite{Molnar:2014fva} the authors model a multicomponent hadron gas with species dependent constant cross sections in order to compute shear viscous phase space corrections. The authors of Ref.~\cite{Denicol:2013nua} compute the hadronic shear viscosity over entropy ratio using different constant cross sections for meson-meson, meson-baryon and baryon-baryon scattering.
\subsection{Massless particles and constant isotropic cross sections}
\label{sec:ResultsA} 
\begin{figure}[t]
\centering
\includegraphics[width=0.6\columnwidth]{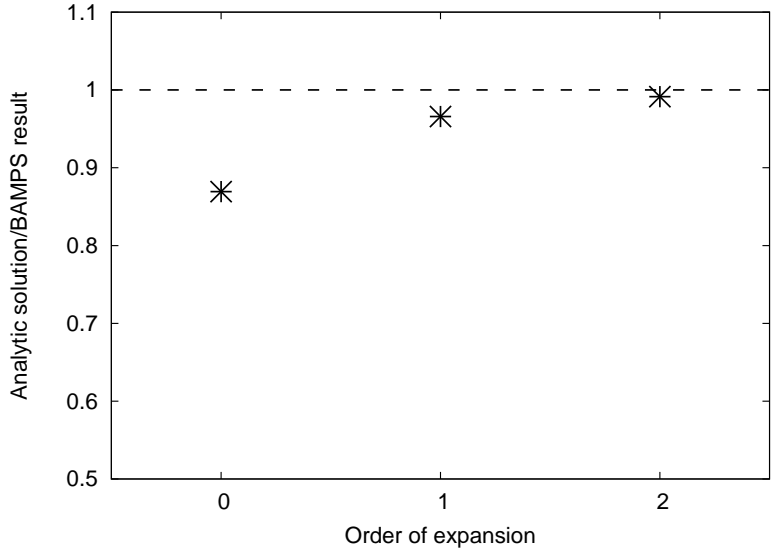} .
\caption{Convergence of the analytical computation for the electric
conductivity for a massless quark-gluon gas towards the numerical value
obtained by the partonic cascade BAMPS (\protect\cite{Greif2014})}
\label{BampsvsAnalytics}
\end{figure}

As a first step, we compute the electric conductivity
for a massless gas of charged and uncharged particles, colliding with a
fixed value of the cross section $\sigma _{\text{tot}}$, which is assumed to
be constant. We give the result for the matrix in Eq.~\eqref{eq:final_matrix}%
, which we truncate at $n=2$. We define $\bar{n}_{ij}=\left( \delta
_{ij}n_{i}n_{T}-n_{i}n_{j}\right) $, with $n_{T}=\sum_{i}^{N_{\text{species}%
}}n_{i}$ being the total particle density. The matrix is 
\begin{equation*}
\mathcal{N}_{mn}^{ij}=%
\begin{pmatrix}
\begin{tabular}{ccc}
$\mathcal{N}_{00}^{ij}$ & $\mathcal{N}_{10}^{ij}$ & $\mathcal{N}_{12}^{ij}$
\\ 
$\mathcal{N}_{10}^{ij}$ & $\mathcal{N}_{11}^{ij}$ & $\mathcal{N}_{12}^{ij}$
\\ 
$\mathcal{N}_{20}^{ij}$ & $\mathcal{N}_{21}^{ij}$ & $\mathcal{N}_{22}^{ij}$
\\ 
&  & 
\end{tabular}%
\end{pmatrix}%
=\sigma _{\text{tot}}%
\begin{pmatrix}
\begin{tabular}{ccc}
$\frac{15}{2}T^{2}\bar{n}_{ij}$ & $36\,T^{3}\bar{n}_{ij}$ & $210\,T^{4}\bar{n%
}_{ij}$ \\ 
$36\,T^{3}\bar{n}_{ij}$ & $T^{4}\left( 216\,\delta
_{ij}n_{i}n_{T}-192\,n_{i}n_{j}\right) $ & $T^{5}\left( 1520\,\delta
_{ij}n_{i}n_{T}-1240\,n_{i}n_{j}\right) $ \\ 
$210\,T^{4}\bar{n}_{ij}$ & $T^{5}\left( 1520\,\delta
_{ij}n_{i}n_{T}-1240\,n_{i}n_{j}\right) $ & $T^{6}\left( 12510\,\delta
_{ij}n_{i}n_{T}-8850\,n_{i}n_{j}\right) $ \\ 
&  & 
\end{tabular}%
\end{pmatrix}%
.
\end{equation*}%
This is the key information to obtain the electric conductivity at order $%
0+1+2$ in the above energy expansion for arbitrary many massless particle
species. In order to compare with previously published numerical solutions
of the BE, we give the explicit result for a gas of seven species, with
electric charges (in units of $e$) $%
q_{1,3}=1/3,q_{2,4}=-1/3,q_{5}=2/3,q_{6}=-2/3,q_{7}=0$ and degeneracys $%
g_{1,2,3,4,5,6}=6,g_{7}=16$, which mimic a quark-gluon plasma. Using that $%
e^{2}=4\pi /137$, and considering a cross section of $\sigma _{\text{tot}}=3~%
\mathrm{mb}$, we obtain the following value of conductivity for this system, 
\begin{equation}
\sigma _{\text{el}}\ =\frac{0.000832737~\mathrm{GeV}^{2}}{T}.
\label{eq:massless_result}
\end{equation}%
In Ref.~\cite{Greif2014} the ultrarelativistic BE was solved for exactly
this configuration (using the partonic cascade BAMPS), and the result
matches the analytic computation of this paper, Eq.~%
\eqref{eq:massless_result}, by about $99\%$. By changing the order of the
expansion, we show in Fig.~\ref{BampsvsAnalytics}, that the result converges
for the considered order in expansion (truncation of the sum in Eq.~%
\eqref{eq:expansion} at $n=2$). 


\subsection{Influence of masses to the electric conductivity}
\label{sec:ResultsB} 
\begin{figure}[t]
\centering
\includegraphics[width=0.6%
\columnwidth]{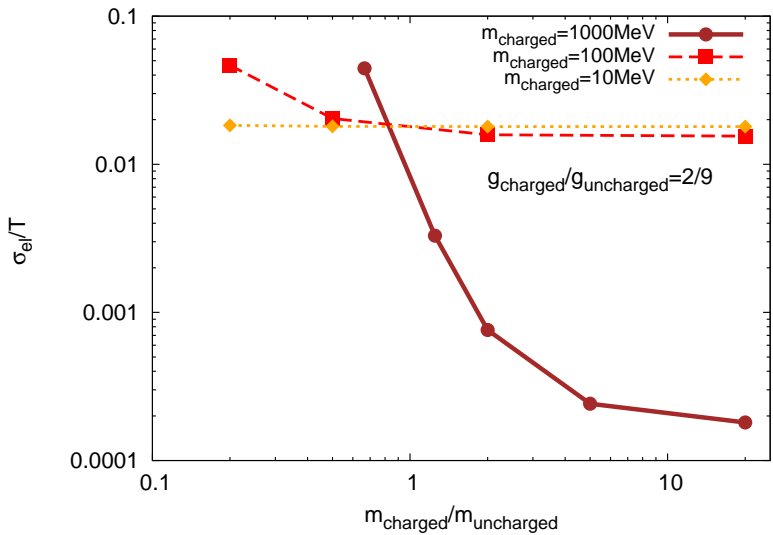}
\caption{The mass dependence of the electric conductivity for (three species
of) interacting relativistic particles. The cross section was arbitrarily
fixed to $10~\mathrm{mb}$, and the degeneracy ratio of charged/uncharged
species is $2/9$, the charges are $\pm 1,0$. On the x-axis we vary the mass
ratio of the charged to the uncharged species, and show results for
different masses of the charged species. (color online)}
\label{fig:MassDependence}
\end{figure}

In order to see the influence of sizeable masses to the
electric conductivity, we consider an arbitrary, simplified scenario for
illustrative purposes. There are three species present, one species with
charge $+1$, degeneracy $1$, one with charge $-1$, degeneracy $1$, and one
with zero charge and degeneracy $9$. All particles have masses, and we vary
the ratio of the mass of the charged species with respect to the mass of the
uncharged species. In Fig.~\ref{fig:MassDependence} we show the results for
the electric conductivity over temperature depending on this ratio, for
different absolute values of the mass of the charged species. There we fix
the cross section to an arbitrary value ($10~\mathrm{mb}$) and set the
temperature to be $140~\mathrm{MeV}$ and the chemical potential is $\mu_q=0$. This is a useful exercise to illustrate the mass dependence.
In thermal and chemical equilibrium, lower mass particles are more abundant
than higher mass particles, and one sees clearly the dependence of the
electric conductivity to the number-density ratio of charged to uncharged
particles. The electric conductivity is clearly very dependent on both the
mass (or density) ratio of charged/uncharged species, and also on the mass
(or density) of the charge carrying species. However, the precise values
need to be computed (finally via numerical integration) as explained in Sec.~
\ref{sec:formalsism_ec}.

\subsection{Pion Gas}

\label{sec:ResultsC} Pions are the most abundant hadrons in an equilibrated
hadron gas. Therefore a pure pion gas can be considered a good starting
point to understand some features of a realistic hadron gas. We set the chemical potential to zero for simplicity. Mainly, pions
interact via the formation and decay of a $\rho $-resonance (see App.~\ref{appsec:resonances}). In Fig.~\ref
{fig:pure_pion} we give results for 3 pion species $\pi ^{+},\pi ^{-},\pi
^{0}$, interacting via Mandelstam s-dependent (isotropic) resonance cross
sections, where we include the dominant $\rho $-meson peak. 
Clearly, the electric conductivity approaches a minimum below $\sim 180~%
\mathrm{MeV}$. This can be physically motivated, as transport coefficients
like the conductivity are expected to show a minimum in the QGP-hadron
crossover region. This region is now believed to be in the vicinity of $\sim
154~\mathrm{MeV}$ \cite{Bazavov:2011nk}.

In Fig.~\ref{fig:pure_pion} we compare our results with the results from
different groups. The brown dash-dotted line represents calculations using
Chiral Perturbation Theory (ChPT) \cite{Fernandez-Fraile2006} and include
only pions. The ChpT-based analysis uses the Green-Kubo formula to extract
the conductivity from the spectral function, identifying the dominant
diagrams in a low energy and low temperature expansion and
implementing unitary of the partial waves in the thermal width. The temperature
dependence of the results from ChPT is very similar to those found in our
results, although the overall magnitude of our electric conductivity is about
a factor of $\sim 1.6$ higher. The blue open diamonds are results obtained
from lattice QCD calculation for an 2+1d anisotropic and unquenched lattice, 
Ref.~\cite{Aarts:2014JHEP}. However, the authors discuss that the lattice data
below the phase transition misses contributions from hadronic interactions,
and should be treated with caution. The grey dashed line is the result
obtained in a conformal Super-Yang Mills plasma \cite{Starinets2006}. In
Ref.~\cite{Finazzo2014, Noronha15}, the authors used a non-conformal, bottom
up holographic model to compute the electric conductivity (cyan dotted
line). The full orange diamonds are results from the pQCD-based partonic
cascade BAMPS ~\cite{Greif2014}, employing a running coupling, leading order, Debye-screened 
pQCD interactions including elastic and inelastic (radiative) scattering of gluons, up, down and
strange quarks. 

\subsection{Pion-Kaon-Nucleon Gas with constant cross sections}
\label{sec:ResultsD} 
\begin{figure}[h!]
\centering
\subfigure[\ Interacting pion gas (red squares) in equilibrium for different temperatures. The pions interact via the $\rho$-resonance-scattering.\label{fig:pure_pion}]{
\includegraphics[width=0.45\columnwidth]{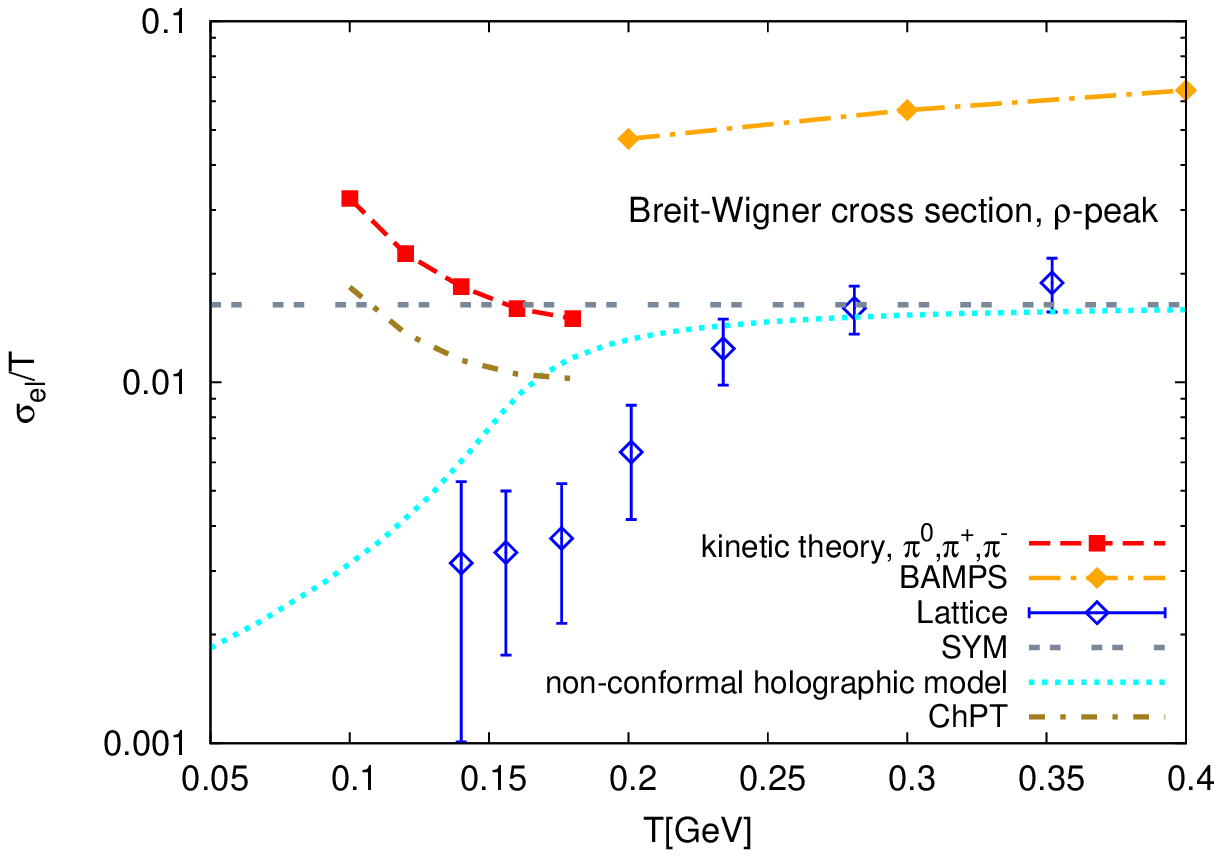}
} 
\subfigure[\ The electric conductivity for interacting pions ($\pi^+,\pi^-,\pi^0$), Kaons ($K^0,\bar{K}^0,K^+,K^-$) and nucleons ($p,\bar{p},n,\bar{n}$). The cross section $\sigma_{\text{tot}}$ is constant and isotropic, and we show results for 4 different values. Results for a pure pion gas are shown for comparison. \label{PionNukleonKaonConstCSGas}]{
\includegraphics[width=0.45\columnwidth]{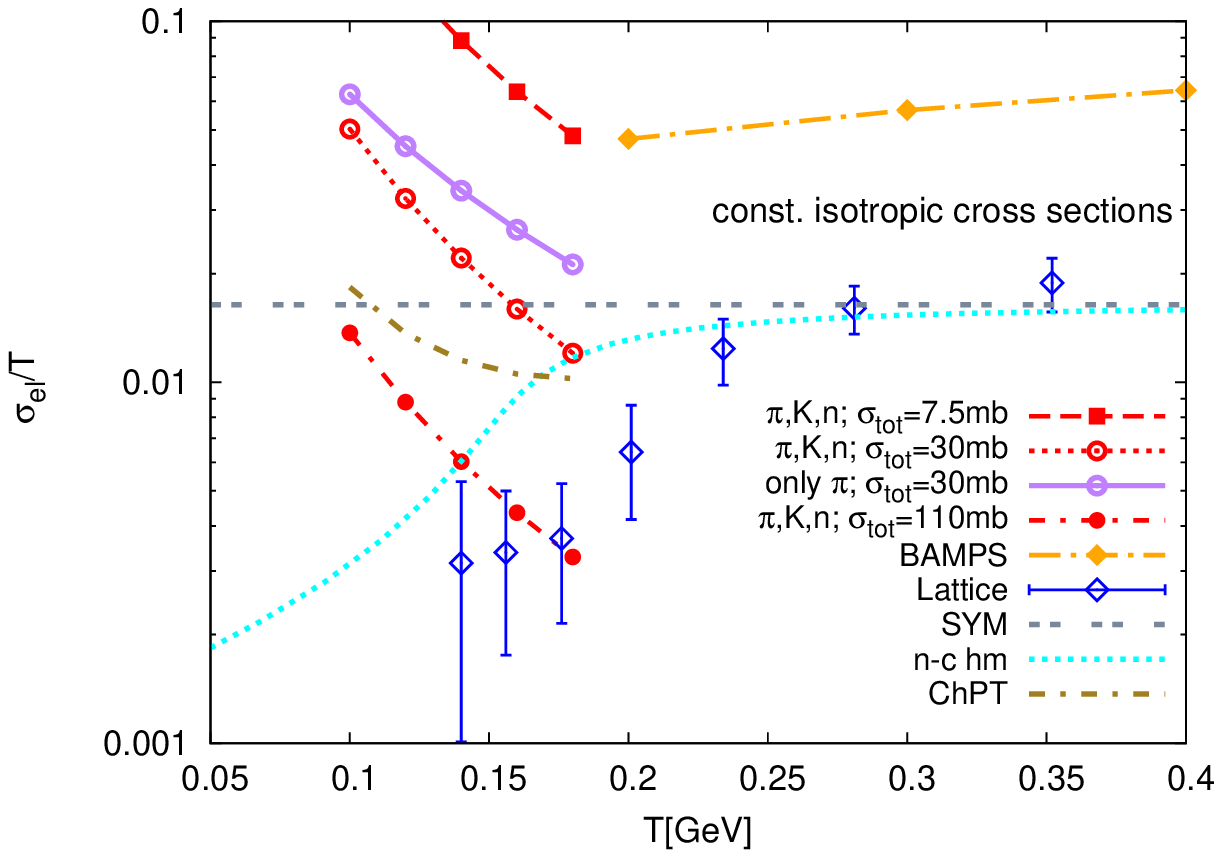}
}
\caption{Results for the electric conductivity from this work and other
theories. Parton transport BAMPS \protect\cite{Greif2014}, Chiral
Perturbation Theory ChPT \protect\cite{Fernandez-Fraile2006}, SYM theory 
\protect\cite{Starinets2006}, a non-conformal holographic model (n-c hm) 
\protect\cite{Finazzo2014} and lattice \citep{Amato2013a, Aarts:2014JHEP}
calculations are shown for comparison. These theories all require very
different effective cross sections when compared to kinetic theory. (color
online)}
\end{figure}

Constant isotropic cross sections are often used to
compare different models or theories. In Fig.~\ref{PionNukleonKaonConstCSGas}
we show results for the electric conductivity for a gas of pions ($\pi
^{+},\pi ^{-},\pi ^{0}$; $m=138~\mathrm{MeV}$), Kaons ($K^{0},\bar{K}%
^{0},K^{+},K^{-}$; $m=496~\mathrm{MeV}$) and nucleons ($p,\bar{p},n,\bar{n}$%
; $m=938~\mathrm{MeV}$), all interacting with a constant cross section $%
\sigma _{\text{tot}}$. The chemical potential is again zero.
We tune this cross section, in order to meet other
calculations at the transition temperature from hadrons to the QGP. Strongly
coupled theories and 2+1d non-quenched lattice require cross section values
of $30-110~\mathrm{mb}$, whereas the pQCD-based partonic cascade BAMPS needs
a value of $\sim 7.5\mathrm{mb}$. These numbers should be taken with care,
as we are dealing here with an oversimplified scenario of effective average
cross sections. Especially as one approaches the crossover region, this
concept is questionable, however it allows to gain some understanding about
the effective coupling strength of different theories. In Fig.\ref%
{PionNukleonKaonConstCSGas}, the full purple line includes only pions, and
uses $\sigma _{\text{tot}}=30~\mathrm{mb}$. By comparing with the dashed red
line (all species), one sees the influence of other, heavier species. We
expect, that the inclusion of even more species, albeit not very abundant,
may decrease the electric conductivity. This may be true even in the case of
realistic $s$-dependent cross sections, cf. Sec.~\ref{sec:ResultsE}. 

\subsection{Pion-Kaon-Nucleon Gas with experimental cross sections}
\label{sec:ResultsE}
\begin{figure}[t!]
\centering
\includegraphics[width=0.6\columnwidth]{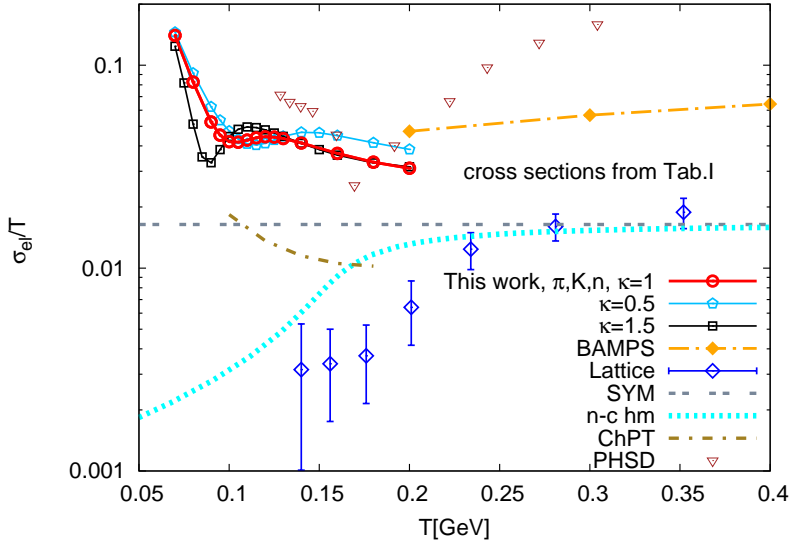}
\caption{Results for the electric conductivity from this work, including pions, kaons and nucleons, compared to results from PHSD \cite{Cassing2013,Steinert2013} and all other theories as before. All constant cross sections (see Tab.~\ref{tab1}) are multiplied with a factor $\kappa$, which we change in the range of $\kappa=0.5,1,1.5$ in order to get a feeling for the uncertainty. (color online)}
\label{fig:full_case}
\end{figure}
The calculation procedure presented in this paper becomes gradually more complicated as more particle species are included, with the final numerical integrations becoming rather tedious and time consuming.  Furthermore, all cross sections among all species have to be known, something quite problematic in the hadronic zoo. In order to get a rough picture of the electric conductivity in a hadron gas, we use pions, kaons and nucleons as in the previous section, but include now as realistic cross sections as possible, as shown in Tab.\ref{tab1}. Many of them are approximated by constant values, but we include different resonances. The chemical potential is zero. The result is shown in Fig.~\ref{fig:full_case}. In order to get a handle on the uncertainty introduced by using approximated constant cross sections $\sigma_{\text{const.}}$ we multiply these with a factor $\kappa$, $\sigma_{\text{const.}}\rightarrow\sigma_{\text{const.}}\kappa $, and vary $\kappa=0.5,1,1.5$. The change of the conductivity is visible but not dramatic.
Due to the presence of charged kaons (and nucleons, these are not as important), the conductivity is now higher ($30-130\%$) compared to the pure pion case from Fig.~\ref{fig:pure_pion}. The dip in the conductivity at around $T=100~\mathrm{MeV}$ is prominent. To explain this feature, we note first that the mean invariant collision energy $\llkl\sqrt{s}\rrkl$ from the $2\leftrightarrow 2$ collisions, whose effect we are effectively studying, is temperature dependent.
For some combinations of species and temperatures, $\llkl\sqrt{s}\rrkl$ lies in the region of a resonance peak, thus the electric conductivity decreases. However, one has to be cautious, as this dip in the conductivity will disappear or shift for another choice of the set of cross sections, or the inclusion of more resonances. This can be seen by the fact, that the dip moves as we vary $\kappa$. 
 As can be seen by comparing with Fig.~\ref{PionNukleonKaonConstCSGas}, the overall magnitude in our case is dominated by the constant cross section values, mostly $\sim10~\mathrm{mb}$ in the important channels. 
Although results have to be taken with care due to the uncertainties in the cross sections, they are (to our knowledge) the first (semi-)analytic kinetic computation of the electric conductivity in the hadronic sector for multiple species. 

In Fig.~\ref{fig:full_case} we also compare to results from the Parton-Hadron-String Dynamics (PHSD) approach \cite{Cassing2013,Steinert2013}. PHSD is in the hadronic sector a covariant extension to the Boltzmann-Uehling-Uhlenbeck model \cite{Cassing:1990dr}. The authors apply an electric current to the numerical simulation in thermal equilibrium and observe a static current in order to extract the electric conductivity. The hadronic sector contains several mesons and baryons with resonance cross sections. Their results (triangles) are in the same order of magnitude as ours, and closer than the results from other groups.

\begin{table}[t!]
\begin{tabular}{|c||c|c|c|c|c|c|c|c|c|c|c|}
\hline 
\rule[-1ex]{0pt}{2.5ex} $\quad$ &$\pi^+$ & $\pi^-$  & $\pi^0$  & $K^+$  & $K^-$ & $K^0$  & $\bar{K}^0$  & $p$ & $n$ & $\bar{p}$ & $\bar{n}$ \\ 
\hhline{|=||=|=|=|=|=|=|=|=|=|=|=|}
\rule[-1ex]{0pt}{2.5ex}$\pi^+$ & $10$ & $\rho$  & $\rho$  & $10$  & $10$ & $K^\star$  & $10$  & $\Delta$ & $10$ & $10$ & $\Delta$ \\ 
\hline 
\rule[-1ex]{0pt}{2.5ex}$\pi^-$ & \  & $10$  & $\rho$  & $K^\star$  & $10$ & $10$  & $K^\star$  & $10$ & $\Delta$ & $\Delta$ & $10$ \\ 
\hline 
\rule[-1ex]{0pt}{2.5ex}$\pi^0$ & \  & $$  & $5$  & $K^\star$  & $10$ & $K^\star$  & $K^\star$  & $\Delta$ & $\Delta$ & $\Delta$ & $\Delta$ \\ 
\hline 
\rule[-1ex]{0pt}{2.5ex}$K^+$ & \  & \   & \   & $10$  & $10$ & $10$  & $50$  & $6$ & $10$ & $20$ & $10$ \\ 
\hline 
\rule[-1ex]{0pt}{2.5ex}$K^-$ & \  & \   & \   & \   & $10$ & $50$  & $10$  & $20$ & $10$ & $6$ & $10$ \\ 
\hline 
\rule[-1ex]{0pt}{2.5ex}$K^0$ & \  & \   & \   & \   & \  & $10$  & $50$  & $6$ & $6$ & $20$ & $20$ \\ 
\hline 
\rule[-1ex]{0pt}{2.5ex}$\bar{K}^0$ & \  & \   & \   & \   & \  & \   & $10$  & $8$ & $20$ & $6$ & $6$ \\ 
\hline 
\rule[-1ex]{0pt}{2.5ex}$p$ & \  & \   & \   & \   & \  & \   & \   & $20$ & $20$ & $100$ & $20$ \\ 
\hline 
\rule[-1ex]{0pt}{2.5ex}$n$ & \  & \   & \   & \   & \  & \   & \   & \  & $20$ & $20$ & $100$ \\ 
\hline 
\rule[-1ex]{0pt}{2.5ex}$\bar{p}$ & \  & \   & \   & \   & \  & \   & \   & \  & \  & $10$ & $10$ \\ 
\hline 
\rule[-1ex]{0pt}{2.5ex}$\bar{n}$ & \  & \   & \   & \   & \  & \   & \   & \  & \  & \  & $10$ \\ 
\hline 
\end{tabular}
\caption{The cross sections we used among all species. Numbers are in mb, $\rho,K^\star$ and $\Delta$ denote Breit-Wigner shaped cross sections with those resonances. Values taken from \cite{UrQMD1,UrQMD2,GiBUU,PDG}. Complicated or unknown functional forms are approximated by an average constant value. The results depend modestly on the choice of these parametrisations, compare also the results from Sec.\ref{sec:ResultsD})} 
\label{tab1}
\end{table}

\section{Conclusions}
\label{sec:conclusion}
In summary, we have developed an analytic formalism to compute the electric conductivity of relativistic, massless or massive gases, governed by the linearized Boltzmann equation including elastic scattering. 
We use the full linearized collision term, and are able to include arbitrary cross section parametrizations. Naturally, all species in a thermal medium can interact with each other and charged species contribute to the electric current, whereas uncharged species act as a resistance for the current. 
The formalism can be reduced to the well-known Drude formula for the electric conductivity. It involves a complicated matrix inversion, and, to be exact, the computation of infinitely many kinetic integrals. For massive species, these have to be evaluated using numerical integration methods. The expansion we make converges rather fast. By comparing to previously published numerical results for the massless case we find excellent agreement. The formalism is quite general and can be extendend in various ways, thus we start by investigating the dependence of the conductivity to masses and mass ratios of charged to uncharged species. Ultimately, we use the formalism to present results for the electric conductivity of a massive pion gas, including all pion species, and experimentally measured (Breit-Wigner resonance) cross-sections. 
We see in accordance with other published results, that the conductivity decreases with increasing temperature, approaching results from a non-conformal holographic model. The temperature dependence is very similar to results from ChPT. 
Furthermore we include pions, kaons and nucleons with their masses, and present results for a constant isotropic cross section. In this simplified case we can obtain effective cross sections in the range from $\sim 7-100~\mathrm{mb}$ when compared to other theories as pQCD parton transport or lattice.  
We extend the study further, and present results using a set of approximated realistic cross sections, including resonances. In the present paper, we restrict the results to zero chemical potential. The influence of chemical potentials will be adressed in future. 
Clearly, the limiting factor is the lack of precise knowledge of elastic cross sections among the hadrons, and our results depend on the choice of their parametrisation. Unknown or complicated cross sections can only be approximated by energy independent constants. However, we believe that the inclusion of fairly realistic cross sections involving pions and kaons, including the $\rho,\Delta$ and $K^\star$ resonance renders the result physical. The cross sections among protons or neutrons play only a minor role for the final result, as these particles are less abundant due to their mass\footnote{As a note, only processes involving electrically charged particles contribute, e.g. the neutron-neutron cross section is actually irrelevant.}. The approximation we made by neglecting heavier particles is thus well justified, however, in future, the study can naturally be extended to include more particle species and their resonances. It is also possible to compute other transport coefficients in a similar fashion.

\section{Acknowledgments}
MG is grateful to ``Helmhotz
Graduate School for Heavy Ion research''. GSD is supported under DOE Contract No. DE- SC0012704. The authors are grateful to the Center for Scientific Computing (CSC) Frankfurt for the computing resources. This work was supported by the Helmholtz International Center for FAIR within the framework of the LOEWE program launched by the State of Hesse.

\appendix
\section{Calculations of the collision integrals}
\label{appsec:matrix_element_examples}
In the calculations of the matrix elements, the following integrals have to be solved. We will show only some examples, all other integrals can be worked out in a similar fashion.
Consider the following integral,
\begin{align}
\int \mathrm{d}P \mathrm{d}P^\prime (2\pi)^6 s \sigma_{ij}(s,\Theta)  \delta^{(4)}(k^\mu + k^{\prime\mu}-p^\mu - p^{\prime\mu}) p^\alpha \equiv \Gamma^\alpha.
\end{align} 
We define a unitless vector (normalized total momentum of the collision) $\tilde{P}_T^\mu=(k^\mu+k^{\prime\mu})/\sqrt{s}$, and the projection orthogonal to it, $\Delta_P^{\mu\nu}=g^{\mu\nu}-\tilde{P}_T^\mu\tilde{P}_T^\nu$. The tensor $\Gamma^\alpha$ can only depend on  $\tilde{P}_T^\mu$, so we can decompose,
\begin{align}
\Gamma^\alpha = a(s) \tilde{P}_T^\alpha,\quad a(s) =\tilde{P}_T^\alpha \Gamma_\alpha
\end{align}
where
\begin{align}
a_{ij}(s)=\gamma_{ij}\int \mathrm{d}P \mathrm{d}P^\prime (2\pi)^6 s \sigma_{ij}(s,\Theta)  \delta^{(4)}(k_i^\mu + k_j^{\prime\mu}-p_i^\mu - p_j^{\prime\mu}) (p_i^\alpha \tilde{P}_{T,\alpha}).
\end{align}
We can always evaluate a scalar integral in the center of momentum/center of mass frame, where $p_i^\alpha \tilde{P}_{T,\alpha}=p_i^0$.
In the massless case, $a=\sigma_{ij}(s,\Theta) s\sqrt{s}/4$, in the massive case,
\begin{align}
a_{ij}(s)&=\gamma_{ij}\int\frac{\mathrm{d}^3p}{p^0_i}\frac{\mathrm{d}^3p}{p^0_i} s\sigma_{ij}(s,\Theta) \delta(p^0_i + p^{\prime 0}_j-\sqrt{s})\delta^{(3)}(\textbf{p}_i+\textbf{p}_j^\prime)p^0_i \n
&= \gamma_{ij}\int\frac{|\textbf{r}|^2\mathrm{d}|\textbf{r}|}{p^0_ip^{\prime 0}_j} s\sigma_{ij}(s,\Theta) \delta(p^0_i + p^{\prime 0}_j-\sqrt{s})\delta^{(3)}(\textbf{p}_i+\textbf{p}_j^\prime)p^0_i \n
&= \frac{1}{2}\kl\gamma_{ij}\int\mathrm{d}\Omega \sigma_{ij}(s,\Theta)\kr\sqrt{(s-s_a^{ij})(s-s_b^{ij})}\sqrt{\frac{1}{4s}(s-s_a^{ij})(s-s_b^{ij})+m_i^2}
\end{align}
where we defined 
\begin{align}
|\textbf{r}|=\frac{1}{2x}\sqrt{\kl x^2-(m_i+m_j)^2 \kr \kl x^2-(m_i-m_j)^2 \kr},\quad s_a^{ij}=(m_i+m_j)^2,\quad s_b^{ij}=(m_i-m_j)^2,\quad x=p^0_i + p^0_j,\quad  
\end{align}
and use
\begin{align}
\frac{\mathrm{d}x}{x}=\frac{|\textbf{r}|\mathrm{d}|\textbf{r}|}{p^0_i p^0_j}.
\end{align}
The $\mathrm{d}K\mathrm{d}K^\prime$-integrals of Eq.~\eqref{eq:integrals} are easily done in the massless case, but require numerical integration in the massive case.

\section{Cross-sections for pion-Isotriplett elastic scattering via $\rho$ resonances}
\label{appsec:resonances}
As an example for the resonance cross sections, the total cross-section for the reaction
\begin{equation}
\pi^{\pm}+\pi^{\mp}\rightarrow \rho^0 \rightarrow \pi^{\pm} + \pi^{\mp}
\label{decay1}
\end{equation}
is given by (we use the parametrisation given e.g. in \cite{UrQMD1,UrQMD2})
\begin{equation}
\sigma_{\text{tot}}(\sqs)=\llkl j_{\pi^{\mp}},m_{\pi^{\mp}},j_{\pi^{\pm}},m_{\pi^{\pm}}\vert\vert J_{\rho^0},M_{\rho^0}\rrkl \frac{2S_{\rho^0}+1}{(2S_{\pi^{\mp}}+1)(2S_{\pi^{\pm}}+1)}\frac{\pi}{p^2_{\text{CMS}}}\frac{\Gamma_{\rho^0 \rightarrow \pi^{\pm} + \pi^{\mp}}\Gamma_{\text{tot}}}{(M_{\rho^0}-\sqs)^2+\frac{\Gamma^2_{\text{tot}}}{4}}
\end{equation}
Here, $j,J$ is the isospin of the particle or resonance, $S_{\text{particle}}$ its spin and $m,M$ the z-component of it.
The Clebsch-Gordon coefficients can be looked up:\\
\begin{center}
 \begin{tabular}{|c|c|}
\hline 
$\llkl j_{\pi^{\mp}},m_{\pi^{\mp}},j_{\pi^{\pm}},m_{\pi^{\pm}}\vert\vert J_{\rho^0},M_{\rho^0}\rrkl$ & $\mp \sqrt{\frac{1}{2}}$ \\ 
\hline
$\llkl j_{\pi^{-}},m_{\pi^{-}},j_{\pi^{0}},m_{\pi^{0}}\vert\vert J_{\rho^-},M_{\rho^-}\rrkl$ & $-\frac{1}{2}$ \\ 
\hline 
$\llkl j_{\pi^{0}},m_{\pi^{0}},j_{\pi^{-}},m_{\pi^{-}}\vert\vert J_{\rho^-},M_{\rho^-}\rrkl$ & $\frac{1}{2}$ \\ 
\hline 
$\llkl j_{\pi^{+}},m_{\pi^{+}},j_{\pi^{0}},m_{\pi^{0}}\vert\vert J_{\rho^+},M_{\rho^+}\rrkl$ & $\frac{1}{2}$ \\ 
\hline 
$\llkl j_{\pi^{0}},m_{\pi^{0}},j_{\pi^{-}},m_{\pi^{-}}\vert\vert J_{\rho^-},M_{\rho^-}\rrkl$ & $-\frac{1}{2}$ \\ 
\hline 
\end{tabular}
 \end{center} 
The Center-of-Mass momentum is given by
\begin{equation}
p_{\text{CMS}}=\frac{1}{2\sqrt{s}}\sqrt{(s-(m_{\pi^+}+m_{\pi^-})^2)\m(s-(m_{\pi^+}-m_{\pi^-})^2) }.
\end{equation}
The widths are themselves energy-dependent:
\begin{align}
\Gamma_{\rho^0 \rightarrow \pi^{\pm} + \pi^{\mp}}(\sqrt{s})=\Gamma^{\text{pole}}_{\rho^0 \rightarrow \pi^{\pm}+ \pi^{\mp}} \frac{m_\rho}{\sqrt{s}}\kl\frac{p_\text{CMS}(\sqrt{s})}{p_\text{CMS}(m_\rho)}\kr^{2l+1}\frac{1.2}{1+0.2 \kl\frac{p_\text{CMS}(\sqrt{s})}{p_\text{CMS}(m_\rho)}\kr^{2l}},
\end{align}
with an angular momentum $l$ of the decay.
We are considering only one decay channel for each process, so $\Gamma_\text{tot}=\Gamma_{\text{decay channel}}$.
\bibliographystyle{apsrev4-1}
\bibliography{library_manuell.bib}

\begin{thebibliography}{68}%
\makeatletter
\providecommand \@ifxundefined [1]{%
 \@ifx{#1\undefined}
}%
\providecommand \@ifnum [1]{%
 \ifnum #1\expandafter \@firstoftwo
 \else \expandafter \@secondoftwo
 \fi
}%
\providecommand \@ifx [1]{%
 \ifx #1\expandafter \@firstoftwo
 \else \expandafter \@secondoftwo
 \fi
}%
\providecommand \natexlab [1]{#1}%
\providecommand \enquote  [1]{``#1''}%
\providecommand \bibnamefont  [1]{#1}%
\providecommand \bibfnamefont [1]{#1}%
\providecommand \citenamefont [1]{#1}%
\providecommand \href@noop [0]{\@secondoftwo}%
\providecommand \href [0]{\begingroup \@sanitize@url \@href}%
\providecommand \@href[1]{\@@startlink{#1}\@@href}%
\providecommand \@@href[1]{\endgroup#1\@@endlink}%
\providecommand \@sanitize@url [0]{\catcode `\\12\catcode `\$12\catcode
  `\&12\catcode `\#12\catcode `\^12\catcode `\_12\catcode `\%12\relax}%
\providecommand \@@startlink[1]{}%
\providecommand \@@endlink[0]{}%
\providecommand \url  [0]{\begingroup\@sanitize@url \@url }%
\providecommand \@url [1]{\endgroup\@href {#1}{\urlprefix }}%
\providecommand \urlprefix  [0]{URL }%
\providecommand \Eprint [0]{\href }%
\providecommand \doibase [0]{http://dx.doi.org/}%
\providecommand \selectlanguage [0]{\@gobble}%
\providecommand \bibinfo  [0]{\@secondoftwo}%
\providecommand \bibfield  [0]{\@secondoftwo}%
\providecommand \translation [1]{[#1]}%
\providecommand \BibitemOpen [0]{}%
\providecommand \bibitemStop [0]{}%
\providecommand \bibitemNoStop [0]{.\EOS\space}%
\providecommand \EOS [0]{\spacefactor3000\relax}%
\providecommand \BibitemShut  [1]{\csname bibitem#1\endcsname}%
\let\auto@bib@innerbib\@empty
\bibitem [{\citenamefont {Arsene}\ \emph {et~al.}(2005)\citenamefont {Arsene}
  \emph {et~al.}}]{Arsene:2004fa}%
  \BibitemOpen
  \bibfield  {author} {\bibinfo {author} {\bibfnamefont {I.}~\bibnamefont
  {Arsene}} \emph {et~al.} (\bibinfo {collaboration} {BRAHMS}),\ }\href
  {\doibase 10.1016/j.nuclphysa.2005.02.130} {\bibfield  {journal} {\bibinfo
  {journal} {Nucl. Phys.}\ }\textbf {\bibinfo {volume} {A757}},\ \bibinfo
  {pages} {1} (\bibinfo {year} {2005})},\ \Eprint
  {http://arxiv.org/abs/nucl-ex/0410020} {arXiv:nucl-ex/0410020} \BibitemShut
  {NoStop}%
\bibitem [{\citenamefont {Adcox}\ \emph {et~al.}(2005)\citenamefont {Adcox}
  \emph {et~al.}}]{Adcox:2004mh}%
  \BibitemOpen
  \bibfield  {author} {\bibinfo {author} {\bibfnamefont {K.}~\bibnamefont
  {Adcox}} \emph {et~al.} (\bibinfo {collaboration} {PHENIX}),\ }\href
  {\doibase 10.1016/j.nuclphysa.2005.03.086} {\bibfield  {journal} {\bibinfo
  {journal} {Nucl. Phys.}\ }\textbf {\bibinfo {volume} {A757}},\ \bibinfo
  {pages} {184} (\bibinfo {year} {2005})},\ \Eprint
  {http://arxiv.org/abs/nucl-ex/0410003} {arXiv:nucl-ex/0410003} \BibitemShut
  {NoStop}%
\bibitem [{\citenamefont {Back}\ \emph {et~al.}(2005)\citenamefont {Back} \emph
  {et~al.}}]{Back:2004je}%
  \BibitemOpen
  \bibfield  {author} {\bibinfo {author} {\bibfnamefont {B.~B.}\ \bibnamefont
  {Back}} \emph {et~al.},\ }\href {\doibase 10.1016/j.nuclphysa.2005.03.084}
  {\bibfield  {journal} {\bibinfo  {journal} {Nucl. Phys.}\ }\textbf {\bibinfo
  {volume} {A757}},\ \bibinfo {pages} {28} (\bibinfo {year} {2005})},\ \Eprint
  {http://arxiv.org/abs/nucl-ex/0410022} {arXiv:nucl-ex/0410022} \BibitemShut
  {NoStop}%
\bibitem [{\citenamefont {Adams}\ \emph {et~al.}(2005)\citenamefont {Adams}
  \emph {et~al.}}]{Adams:2005dq}%
  \BibitemOpen
  \bibfield  {author} {\bibinfo {author} {\bibfnamefont {J.}~\bibnamefont
  {Adams}} \emph {et~al.} (\bibinfo {collaboration} {STAR}),\ }\href {\doibase
  10.1016/j.nuclphysa.2005.03.085} {\bibfield  {journal} {\bibinfo  {journal}
  {Nucl. Phys.}\ }\textbf {\bibinfo {volume} {A757}},\ \bibinfo {pages} {102}
  (\bibinfo {year} {2005})},\ \Eprint {http://arxiv.org/abs/nucl-ex/0501009}
  {arXiv:nucl-ex/0501009} \BibitemShut {NoStop}%
\bibitem [{\citenamefont {Kolb}\ \emph {et~al.}(2000)\citenamefont {Kolb},
  \citenamefont {Sollfrank},\ and\ \citenamefont {Heinz}}]{Kolb:2000sd}%
  \BibitemOpen
  \bibfield  {author} {\bibinfo {author} {\bibfnamefont {P.~F.}\ \bibnamefont
  {Kolb}}, \bibinfo {author} {\bibfnamefont {J.}~\bibnamefont {Sollfrank}}, \
  and\ \bibinfo {author} {\bibfnamefont {U.~W.}\ \bibnamefont {Heinz}},\ }\href
  {\doibase 10.1103/PhysRevC.62.054909} {\bibfield  {journal} {\bibinfo
  {journal} {Phys. Rev.}\ }\textbf {\bibinfo {volume} {C62}},\ \bibinfo {pages}
  {054909} (\bibinfo {year} {2000})},\ \Eprint
  {http://arxiv.org/abs/hep-ph/0006129} {arXiv:hep-ph/0006129} \BibitemShut
  {NoStop}%
\bibitem [{\citenamefont {Gale}\ \emph {et~al.}(2013)\citenamefont {Gale},
  \citenamefont {Jeon},\ and\ \citenamefont {Schenke}}]{GALE2013}%
  \BibitemOpen
  \bibfield  {author} {\bibinfo {author} {\bibfnamefont {C.}~\bibnamefont
  {Gale}}, \bibinfo {author} {\bibfnamefont {S.}~\bibnamefont {Jeon}}, \ and\
  \bibinfo {author} {\bibfnamefont {B.}~\bibnamefont {Schenke}},\ }\href
  {\doibase 10.1142/S0217751X13400113} {\bibfield  {journal} {\bibinfo
  {journal} {International Journal of Modern Physics A}\ }\textbf {\bibinfo
  {volume} {28}},\ \bibinfo {pages} {1340011} (\bibinfo {year}
  {2013})}\BibitemShut {NoStop}%
\bibitem [{\citenamefont {Schenke}(2011)}]{Schenke2011}%
  \BibitemOpen
  \bibfield  {author} {\bibinfo {author} {\bibfnamefont {B.}~\bibnamefont
  {Schenke}},\ }\href {http://iopscience.iop.org/0954-3899/38/12/124009}
  {\bibfield  {journal} {\bibinfo  {journal} {Journal of Physics G}\ ,\
  \bibinfo {pages} {1}} (\bibinfo {year} {2011})},\ \Eprint
  {http://arxiv.org/abs/1106.6012v1} {arXiv:1106.6012v1} \BibitemShut {NoStop}%
\bibitem [{\citenamefont {Shen}\ and\ \citenamefont
  {Heinz}(2012)}]{PhysRevC.85.054902}%
  \BibitemOpen
  \bibfield  {author} {\bibinfo {author} {\bibfnamefont {C.}~\bibnamefont
  {Shen}}\ and\ \bibinfo {author} {\bibfnamefont {U.}~\bibnamefont {Heinz}},\
  }\href {\doibase 10.1103/PhysRevC.85.054902} {\bibfield  {journal} {\bibinfo
  {journal} {Phys. Rev. C}\ }\textbf {\bibinfo {volume} {85}},\ \bibinfo
  {pages} {054902} (\bibinfo {year} {2012})}\BibitemShut {NoStop}%
\bibitem [{\citenamefont {Kolb}\ and\ \citenamefont
  {Heinz}(2003)}]{Kolb:2003dz}%
  \BibitemOpen
  \bibfield  {author} {\bibinfo {author} {\bibfnamefont {P.~F.}\ \bibnamefont
  {Kolb}}\ and\ \bibinfo {author} {\bibfnamefont {U.~W.}\ \bibnamefont
  {Heinz}},\ }\href@noop {} {\  (\bibinfo {year} {2003})},\ \Eprint
  {http://arxiv.org/abs/nucl-th/0305084} {arXiv:nucl-th/0305084} \BibitemShut
  {NoStop}%
\bibitem [{\citenamefont {Teaney}\ \emph {et~al.}(2001)\citenamefont {Teaney},
  \citenamefont {Lauret},\ and\ \citenamefont {Shuryak}}]{Teaney:2001av}%
  \BibitemOpen
  \bibfield  {author} {\bibinfo {author} {\bibfnamefont {D.}~\bibnamefont
  {Teaney}}, \bibinfo {author} {\bibfnamefont {J.}~\bibnamefont {Lauret}}, \
  and\ \bibinfo {author} {\bibfnamefont {E.~V.}\ \bibnamefont {Shuryak}},\
  }\href {\doibase 10.1103/PhysRevLett.86.4783} {\bibfield  {journal} {\bibinfo
   {journal} {Phys. Rev. Lett.}\ }\textbf {\bibinfo {volume} {86}},\ \bibinfo
  {pages} {4783} (\bibinfo {year} {2001})}\BibitemShut {NoStop}%
\bibitem [{\citenamefont {Del~Zanna}\ \emph {et~al.}(2013)\citenamefont
  {Del~Zanna}, \citenamefont {Chandra}, \citenamefont {Inghirami},
  \citenamefont {Rolando}, \citenamefont {Beraudo}, \citenamefont {De~Pace},
  \citenamefont {Pagliara}, \citenamefont {Drago},\ and\ \citenamefont
  {Becattini}}]{DelZanna:2013eua}%
  \BibitemOpen
  \bibfield  {author} {\bibinfo {author} {\bibfnamefont {L.}~\bibnamefont
  {Del~Zanna}}, \bibinfo {author} {\bibfnamefont {V.}~\bibnamefont {Chandra}},
  \bibinfo {author} {\bibfnamefont {G.}~\bibnamefont {Inghirami}}, \bibinfo
  {author} {\bibfnamefont {V.}~\bibnamefont {Rolando}}, \bibinfo {author}
  {\bibfnamefont {A.}~\bibnamefont {Beraudo}}, \bibinfo {author} {\bibfnamefont
  {A.}~\bibnamefont {De~Pace}}, \bibinfo {author} {\bibfnamefont
  {G.}~\bibnamefont {Pagliara}}, \bibinfo {author} {\bibfnamefont
  {A.}~\bibnamefont {Drago}}, \ and\ \bibinfo {author} {\bibfnamefont
  {F.}~\bibnamefont {Becattini}},\ }\href {\doibase
  10.1140/epjc/s10052-013-2524-5} {\bibfield  {journal} {\bibinfo  {journal}
  {Eur. Phys. J.}\ }\textbf {\bibinfo {volume} {C73}},\ \bibinfo {pages} {2524}
  (\bibinfo {year} {2013})},\ \Eprint {http://arxiv.org/abs/1305.7052}
  {arXiv:1305.7052} \BibitemShut {NoStop}%
\bibitem [{\citenamefont {Karpenko}\ \emph {et~al.}(2014)\citenamefont
  {Karpenko}, \citenamefont {Huovinen},\ and\ \citenamefont
  {Bleicher}}]{Karpenko:2013wva}%
  \BibitemOpen
  \bibfield  {author} {\bibinfo {author} {\bibfnamefont {I.}~\bibnamefont
  {Karpenko}}, \bibinfo {author} {\bibfnamefont {P.}~\bibnamefont {Huovinen}},
  \ and\ \bibinfo {author} {\bibfnamefont {M.}~\bibnamefont {Bleicher}},\
  }\href {\doibase 10.1016/j.cpc.2014.07.010} {\bibfield  {journal} {\bibinfo
  {journal} {Comput. Phys. Commun.}\ }\textbf {\bibinfo {volume} {185}},\
  \bibinfo {pages} {3016} (\bibinfo {year} {2014})},\ \Eprint
  {http://arxiv.org/abs/1312.4160} {arXiv:1312.4160} \BibitemShut {NoStop}%
\bibitem [{\citenamefont {Holopainen}\ \emph {et~al.}(2011)\citenamefont
  {Holopainen}, \citenamefont {Niemi},\ and\ \citenamefont
  {Eskola}}]{Holopainen:2010gz}%
  \BibitemOpen
  \bibfield  {author} {\bibinfo {author} {\bibfnamefont {H.}~\bibnamefont
  {Holopainen}}, \bibinfo {author} {\bibfnamefont {H.}~\bibnamefont {Niemi}}, \
  and\ \bibinfo {author} {\bibfnamefont {K.~J.}\ \bibnamefont {Eskola}},\
  }\href {\doibase 10.1103/PhysRevC.83.034901} {\bibfield  {journal} {\bibinfo
  {journal} {Phys. Rev.}\ }\textbf {\bibinfo {volume} {C83}},\ \bibinfo {pages}
  {034901} (\bibinfo {year} {2011})},\ \Eprint {http://arxiv.org/abs/1007.0368}
  {arXiv:1007.0368} \BibitemShut {NoStop}%
\bibitem [{\citenamefont {Xu}\ and\ \citenamefont {Greiner}(2005)}]{Xu2005}%
  \BibitemOpen
  \bibfield  {author} {\bibinfo {author} {\bibfnamefont {Z.}~\bibnamefont
  {Xu}}\ and\ \bibinfo {author} {\bibfnamefont {C.}~\bibnamefont {Greiner}},\
  }\href {\doibase 10.1103/PhysRevC.71.064901} {\bibfield  {journal} {\bibinfo
  {journal} {Physical Review C}\ }\textbf {\bibinfo {volume} {71}},\ \bibinfo
  {pages} {064901} (\bibinfo {year} {2005})},\ \Eprint
  {http://arxiv.org/abs/0406278v2} {arXiv:0406278v2} \BibitemShut {NoStop}%
\bibitem [{\citenamefont {Bouras}\ \emph {et~al.}(2010)\citenamefont {Bouras},
  \citenamefont {Moln\'{a}r}, \citenamefont {Niemi}, \citenamefont {Xu},
  \citenamefont {El}, \citenamefont {Fochler}, \citenamefont {Greiner},\ and\
  \citenamefont {Rischke}}]{Bouras2010a}%
  \BibitemOpen
  \bibfield  {author} {\bibinfo {author} {\bibfnamefont {I.}~\bibnamefont
  {Bouras}}, \bibinfo {author} {\bibfnamefont {E.}~\bibnamefont {Moln\'{a}r}},
  \bibinfo {author} {\bibfnamefont {H.}~\bibnamefont {Niemi}}, \bibinfo
  {author} {\bibfnamefont {Z.}~\bibnamefont {Xu}}, \bibinfo {author}
  {\bibfnamefont {A.}~\bibnamefont {El}}, \bibinfo {author} {\bibfnamefont
  {O.}~\bibnamefont {Fochler}}, \bibinfo {author} {\bibfnamefont
  {C.}~\bibnamefont {Greiner}}, \ and\ \bibinfo {author} {\bibfnamefont
  {D.~H.}\ \bibnamefont {Rischke}},\ }\href
  {http://link.aps.org/doi/10.1103/PhysRevC.82.024910} {\bibfield  {journal}
  {\bibinfo  {journal} {Physical Review C}\ }\textbf {\bibinfo {volume} {82}},\
  \bibinfo {pages} {024910} (\bibinfo {year} {2010})}\BibitemShut {NoStop}%
\bibitem [{\citenamefont {Bouras}\ \emph {et~al.}(2012)\citenamefont {Bouras},
  \citenamefont {El}, \citenamefont {Fochler}, \citenamefont {Niemi},
  \citenamefont {Xu},\ and\ \citenamefont {Greiner}}]{Bouras2012}%
  \BibitemOpen
  \bibfield  {author} {\bibinfo {author} {\bibfnamefont {I.}~\bibnamefont
  {Bouras}}, \bibinfo {author} {\bibfnamefont {A.}~\bibnamefont {El}}, \bibinfo
  {author} {\bibfnamefont {O.}~\bibnamefont {Fochler}}, \bibinfo {author}
  {\bibfnamefont {H.}~\bibnamefont {Niemi}}, \bibinfo {author} {\bibfnamefont
  {Z.}~\bibnamefont {Xu}}, \ and\ \bibinfo {author} {\bibfnamefont
  {C.}~\bibnamefont {Greiner}},\ }\href {\doibase
  10.1016/j.physletb.2012.03.040} {\bibfield  {journal} {\bibinfo  {journal}
  {Physics Letters B}\ }\textbf {\bibinfo {volume} {710}},\ \bibinfo {pages}
  {641} (\bibinfo {year} {2012})},\ \Eprint {http://arxiv.org/abs/1201.5005}
  {arXiv:1201.5005} \BibitemShut {NoStop}%
\bibitem [{\citenamefont {Bouras}\ \emph {et~al.}(2009)\citenamefont {Bouras},
  \citenamefont {Moln\'{a}r}, \citenamefont {Niemi}, \citenamefont {Xu},
  \citenamefont {El}, \citenamefont {Fochler}, \citenamefont {Greiner},\ and\
  \citenamefont {Rischke}}]{Bouras2009}%
  \BibitemOpen
  \bibfield  {author} {\bibinfo {author} {\bibfnamefont {I.}~\bibnamefont
  {Bouras}}, \bibinfo {author} {\bibfnamefont {E.}~\bibnamefont {Moln\'{a}r}},
  \bibinfo {author} {\bibfnamefont {H.}~\bibnamefont {Niemi}}, \bibinfo
  {author} {\bibfnamefont {Z.}~\bibnamefont {Xu}}, \bibinfo {author}
  {\bibfnamefont {A.}~\bibnamefont {El}}, \bibinfo {author} {\bibfnamefont
  {O.}~\bibnamefont {Fochler}}, \bibinfo {author} {\bibfnamefont
  {C.}~\bibnamefont {Greiner}}, \ and\ \bibinfo {author} {\bibfnamefont
  {D.~H.}\ \bibnamefont {Rischke}},\ }\href
  {http://link.aps.org/doi/10.1103/PhysRevLett.103.032301} {\bibfield
  {journal} {\bibinfo  {journal} {Physical Review Letters}\ }\textbf {\bibinfo
  {volume} {103}},\ \bibinfo {pages} {032301} (\bibinfo {year}
  {2009})}\BibitemShut {NoStop}%
\bibitem [{\citenamefont {Fochler}\ \emph {et~al.}(2010)\citenamefont
  {Fochler}, \citenamefont {Xu},\ and\ \citenamefont {Greiner}}]{Fochler2010}%
  \BibitemOpen
  \bibfield  {author} {\bibinfo {author} {\bibfnamefont {O.}~\bibnamefont
  {Fochler}}, \bibinfo {author} {\bibfnamefont {Z.}~\bibnamefont {Xu}}, \ and\
  \bibinfo {author} {\bibfnamefont {C.}~\bibnamefont {Greiner}},\ }\href
  {\doibase 10.1103/PhysRevC.82.024907} {\bibfield  {journal} {\bibinfo
  {journal} {Physical Review C}\ }\textbf {\bibinfo {volume} {82}},\ \bibinfo
  {pages} {024907} (\bibinfo {year} {2010})},\ \Eprint
  {http://arxiv.org/abs/1003.4380} {arXiv:1003.4380} \BibitemShut {NoStop}%
\bibitem [{\citenamefont {Fochler}\ \emph {et~al.}(2011)\citenamefont
  {Fochler}, \citenamefont {Xu},\ and\ \citenamefont {Greiner}}]{Fochler2011}%
  \BibitemOpen
  \bibfield  {author} {\bibinfo {author} {\bibfnamefont {O.}~\bibnamefont
  {Fochler}}, \bibinfo {author} {\bibfnamefont {Z.}~\bibnamefont {Xu}}, \ and\
  \bibinfo {author} {\bibfnamefont {C.}~\bibnamefont {Greiner}},\ }\href
  {\doibase 10.1016/j.nuclphysa.2011.02.095} {\bibfield  {journal} {\bibinfo
  {journal} {Nuclear Physics A}\ }\textbf {\bibinfo {volume} {855}},\ \bibinfo
  {pages} {420} (\bibinfo {year} {2011})},\ \Eprint
  {http://arxiv.org/abs/1012.1811} {arXiv:1012.1811} \BibitemShut {NoStop}%
\bibitem [{\citenamefont {Uphoff}\ \emph {et~al.}(2011)\citenamefont {Uphoff},
  \citenamefont {Fochler}, \citenamefont {Xu},\ and\ \citenamefont
  {Greiner}}]{Uphoff2011a}%
  \BibitemOpen
  \bibfield  {author} {\bibinfo {author} {\bibfnamefont {J.}~\bibnamefont
  {Uphoff}}, \bibinfo {author} {\bibfnamefont {O.}~\bibnamefont {Fochler}},
  \bibinfo {author} {\bibfnamefont {Z.}~\bibnamefont {Xu}}, \ and\ \bibinfo
  {author} {\bibfnamefont {C.}~\bibnamefont {Greiner}},\ }\href {\doibase
  10.1103/PhysRevC.84.024908} {\bibfield  {journal} {\bibinfo  {journal}
  {Physical Review C}\ }\textbf {\bibinfo {volume} {84}},\ \bibinfo {pages}
  {024908} (\bibinfo {year} {2011})},\ \Eprint {http://arxiv.org/abs/1104.2295}
  {arXiv:1104.2295} \BibitemShut {NoStop}%
\bibitem [{\citenamefont {Wesp}\ \emph {et~al.}(2011)\citenamefont {Wesp},
  \citenamefont {El}, \citenamefont {Reining}, \citenamefont {Xu},
  \citenamefont {Bouras},\ and\ \citenamefont {Greiner}}]{Wesp2011}%
  \BibitemOpen
  \bibfield  {author} {\bibinfo {author} {\bibfnamefont {C.}~\bibnamefont
  {Wesp}}, \bibinfo {author} {\bibfnamefont {A.}~\bibnamefont {El}}, \bibinfo
  {author} {\bibfnamefont {F.}~\bibnamefont {Reining}}, \bibinfo {author}
  {\bibfnamefont {Z.}~\bibnamefont {Xu}}, \bibinfo {author} {\bibfnamefont
  {I.}~\bibnamefont {Bouras}}, \ and\ \bibinfo {author} {\bibfnamefont
  {C.}~\bibnamefont {Greiner}},\ }\href {\doibase 10.1103/PhysRevC.84.054911}
  {\bibfield  {journal} {\bibinfo  {journal} {Physical Review C}\ }\textbf
  {\bibinfo {volume} {84}},\ \bibinfo {pages} {054911} (\bibinfo {year}
  {2011})},\ \Eprint {http://arxiv.org/abs/1106.4306} {arXiv:1106.4306}
  \BibitemShut {NoStop}%
\bibitem [{\citenamefont {Reining}\ \emph {et~al.}(2012)\citenamefont
  {Reining}, \citenamefont {Bouras}, \citenamefont {El}, \citenamefont {Wesp},
  \citenamefont {Xu},\ and\ \citenamefont {Greiner}}]{Reining2012}%
  \BibitemOpen
  \bibfield  {author} {\bibinfo {author} {\bibfnamefont {F.}~\bibnamefont
  {Reining}}, \bibinfo {author} {\bibfnamefont {I.}~\bibnamefont {Bouras}},
  \bibinfo {author} {\bibfnamefont {A.}~\bibnamefont {El}}, \bibinfo {author}
  {\bibfnamefont {C.}~\bibnamefont {Wesp}}, \bibinfo {author} {\bibfnamefont
  {Z.}~\bibnamefont {Xu}}, \ and\ \bibinfo {author} {\bibfnamefont
  {C.}~\bibnamefont {Greiner}},\ }\href {\doibase 10.1103/PhysRevE.85.026302}
  {\bibfield  {journal} {\bibinfo  {journal} {Physical Review E}\ }\textbf
  {\bibinfo {volume} {85}},\ \bibinfo {pages} {026302} (\bibinfo {year}
  {2012})},\ \Eprint {http://arxiv.org/abs/1106.4210} {arXiv:1106.4210}
  \BibitemShut {NoStop}%
\bibitem [{\citenamefont {Uphoff}\ \emph {et~al.}(2012)\citenamefont {Uphoff},
  \citenamefont {Fochler}, \citenamefont {Xu},\ and\ \citenamefont
  {Greiner}}]{Uphoff2012}%
  \BibitemOpen
  \bibfield  {author} {\bibinfo {author} {\bibfnamefont {J.}~\bibnamefont
  {Uphoff}}, \bibinfo {author} {\bibfnamefont {O.}~\bibnamefont {Fochler}},
  \bibinfo {author} {\bibfnamefont {Z.}~\bibnamefont {Xu}}, \ and\ \bibinfo
  {author} {\bibfnamefont {C.}~\bibnamefont {Greiner}},\ }\href {\doibase
  10.1016/j.physletb.2012.09.069} {\bibfield  {journal} {\bibinfo  {journal}
  {Physics Letters B}\ }\textbf {\bibinfo {volume} {717}},\ \bibinfo {pages}
  {430} (\bibinfo {year} {2012})},\ \Eprint {http://arxiv.org/abs/1205.4945}
  {arXiv:1205.4945} \BibitemShut {NoStop}%
\bibitem [{\citenamefont {Fochler}\ \emph {et~al.}(2013)\citenamefont
  {Fochler}, \citenamefont {Uphoff}, \citenamefont {Xu},\ and\ \citenamefont
  {Greiner}}]{Fochler2013}%
  \BibitemOpen
  \bibfield  {author} {\bibinfo {author} {\bibfnamefont {O.}~\bibnamefont
  {Fochler}}, \bibinfo {author} {\bibfnamefont {J.}~\bibnamefont {Uphoff}},
  \bibinfo {author} {\bibfnamefont {Z.}~\bibnamefont {Xu}}, \ and\ \bibinfo
  {author} {\bibfnamefont {C.}~\bibnamefont {Greiner}},\ }\href {\doibase
  10.1103/PhysRevD.88.014018} {\bibfield  {journal} {\bibinfo  {journal}
  {Physical Review D}\ }\textbf {\bibinfo {volume} {88}},\ \bibinfo {pages}
  {014018} (\bibinfo {year} {2013})},\ \Eprint {http://arxiv.org/abs/1302.5250}
  {arXiv:1302.5250} \BibitemShut {NoStop}%
\bibitem [{\citenamefont {Greif}\ \emph {et~al.}(2013)\citenamefont {Greif},
  \citenamefont {Reining}, \citenamefont {Bouras}, \citenamefont {Denicol},
  \citenamefont {Xu},\ and\ \citenamefont {Greiner}}]{Greif2013}%
  \BibitemOpen
  \bibfield  {author} {\bibinfo {author} {\bibfnamefont {M.}~\bibnamefont
  {Greif}}, \bibinfo {author} {\bibfnamefont {F.}~\bibnamefont {Reining}},
  \bibinfo {author} {\bibfnamefont {I.}~\bibnamefont {Bouras}}, \bibinfo
  {author} {\bibfnamefont {G.}~\bibnamefont {Denicol}}, \bibinfo {author}
  {\bibfnamefont {Z.}~\bibnamefont {Xu}}, \ and\ \bibinfo {author}
  {\bibfnamefont {C.}~\bibnamefont {Greiner}},\ }\href {\doibase
  10.1103/PhysRevE.87.033019} {\bibfield  {journal} {\bibinfo  {journal}
  {Physical Review E}\ }\textbf {\bibinfo {volume} {87}},\ \bibinfo {pages}
  {033019} (\bibinfo {year} {2013})},\ \Eprint {http://arxiv.org/abs/1301.1190}
  {arXiv:1301.1190} \BibitemShut {NoStop}%
\bibitem [{\citenamefont {Senzel}\ \emph {et~al.}(2015)\citenamefont {Senzel},
  \citenamefont {Fochler}, \citenamefont {Uphoff}, \citenamefont {Xu},\ and\
  \citenamefont {Greiner}}]{Senzel2013}%
  \BibitemOpen
  \bibfield  {author} {\bibinfo {author} {\bibfnamefont {F.}~\bibnamefont
  {Senzel}}, \bibinfo {author} {\bibfnamefont {O.}~\bibnamefont {Fochler}},
  \bibinfo {author} {\bibfnamefont {J.}~\bibnamefont {Uphoff}}, \bibinfo
  {author} {\bibfnamefont {Z.}~\bibnamefont {Xu}}, \ and\ \bibinfo {author}
  {\bibfnamefont {C.}~\bibnamefont {Greiner}},\ }\href {\doibase
  10.1088/0954-3899/42/11/115104} {\bibfield  {journal} {\bibinfo  {journal}
  {J. Phys.}\ }\textbf {\bibinfo {volume} {G42}},\ \bibinfo {pages} {115104}
  (\bibinfo {year} {2015})},\ \Eprint {http://arxiv.org/abs/1309.1657}
  {arXiv:1309.1657} \BibitemShut {NoStop}%
\bibitem [{\citenamefont {Uphoff}\ \emph {et~al.}(2013)\citenamefont {Uphoff},
  \citenamefont {Fochler}, \citenamefont {Xu},\ and\ \citenamefont
  {Greiner}}]{Uphoff2013}%
  \BibitemOpen
  \bibfield  {author} {\bibinfo {author} {\bibfnamefont {J.}~\bibnamefont
  {Uphoff}}, \bibinfo {author} {\bibfnamefont {O.}~\bibnamefont {Fochler}},
  \bibinfo {author} {\bibfnamefont {Z.}~\bibnamefont {Xu}}, \ and\ \bibinfo
  {author} {\bibfnamefont {C.}~\bibnamefont {Greiner}},\ }\href {\doibase
  10.1016/j.nuclphysa.2012.12.102} {\bibfield  {journal} {\bibinfo  {journal}
  {Nuclear Physics A}\ }\textbf {\bibinfo {volume} {910-911}},\ \bibinfo
  {pages} {401} (\bibinfo {year} {2013})},\ \Eprint
  {http://arxiv.org/abs/1208.1970} {arXiv:1208.1970} \BibitemShut {NoStop}%
\bibitem [{\citenamefont {Gyulassy}\ and\ \citenamefont
  {McLerran}(2005)}]{Gyulassy2005}%
  \BibitemOpen
  \bibfield  {author} {\bibinfo {author} {\bibfnamefont {M.}~\bibnamefont
  {Gyulassy}}\ and\ \bibinfo {author} {\bibfnamefont {L.}~\bibnamefont
  {McLerran}},\ }\href {\doibase 10.1016/j.nuclphysa.2004.10.034} {\bibfield
  {journal} {\bibinfo  {journal} {Nuclear Physics A}\ }\textbf {\bibinfo
  {volume} {750}},\ \bibinfo {pages} {30} (\bibinfo {year} {2005})}\BibitemShut
  {NoStop}%
\bibitem [{\citenamefont {Schenke}\ \emph {et~al.}(2011)\citenamefont
  {Schenke}, \citenamefont {Jeon},\ and\ \citenamefont {Gale}}]{Schenke2011a}%
  \BibitemOpen
  \bibfield  {author} {\bibinfo {author} {\bibfnamefont {B.}~\bibnamefont
  {Schenke}}, \bibinfo {author} {\bibfnamefont {S.}~\bibnamefont {Jeon}}, \
  and\ \bibinfo {author} {\bibfnamefont {C.}~\bibnamefont {Gale}},\ }\href
  {http://link.aps.org/doi/10.1103/PhysRevLett.106.042301} {\bibfield
  {journal} {\bibinfo  {journal} {Physical Review Letters}\ }\textbf {\bibinfo
  {volume} {106}},\ \bibinfo {pages} {042301} (\bibinfo {year}
  {2011})}\BibitemShut {NoStop}%
\bibitem [{\citenamefont {Plumari}\ \emph {et~al.}(2012)\citenamefont
  {Plumari}, \citenamefont {Puglisi}, \citenamefont {Scardina},\ and\
  \citenamefont {Greco}}]{Plumari2012}%
  \BibitemOpen
  \bibfield  {author} {\bibinfo {author} {\bibfnamefont {S.}~\bibnamefont
  {Plumari}}, \bibinfo {author} {\bibfnamefont {A.}~\bibnamefont {Puglisi}},
  \bibinfo {author} {\bibfnamefont {F.}~\bibnamefont {Scardina}}, \ and\
  \bibinfo {author} {\bibfnamefont {V.}~\bibnamefont {Greco}},\ }\href
  {http://link.aps.org/doi/10.1103/PhysRevC.86.054902} {\bibfield  {journal}
  {\bibinfo  {journal} {Physical Review C}\ }\textbf {\bibinfo {volume} {86}},\
  \bibinfo {pages} {054902} (\bibinfo {year} {2012})}\BibitemShut {NoStop}%
\bibitem [{\citenamefont {Greif}\ \emph {et~al.}(2014)\citenamefont {Greif},
  \citenamefont {Bouras}, \citenamefont {Greiner},\ and\ \citenamefont
  {Xu}}]{Greif2014}%
  \BibitemOpen
  \bibfield  {author} {\bibinfo {author} {\bibfnamefont {M.}~\bibnamefont
  {Greif}}, \bibinfo {author} {\bibfnamefont {I.}~\bibnamefont {Bouras}},
  \bibinfo {author} {\bibfnamefont {C.}~\bibnamefont {Greiner}}, \ and\
  \bibinfo {author} {\bibfnamefont {Z.}~\bibnamefont {Xu}},\ }\href {\doibase
  10.1103/PhysRevD.90.094014} {\bibfield  {journal} {\bibinfo  {journal} {Phys.
  Rev. D}\ }\textbf {\bibinfo {volume} {90}},\ \bibinfo {pages} {094014}
  (\bibinfo {year} {2014})},\ \Eprint {http://arxiv.org/abs/1408.7049}
  {arXiv:1408.7049} \BibitemShut {NoStop}%
\bibitem [{\citenamefont {Puglisi}\ \emph
  {et~al.}(2014{\natexlab{a}})\citenamefont {Puglisi}, \citenamefont
  {Plumari},\ and\ \citenamefont {Greco}}]{Puglisi2014a}%
  \BibitemOpen
  \bibfield  {author} {\bibinfo {author} {\bibfnamefont {A.}~\bibnamefont
  {Puglisi}}, \bibinfo {author} {\bibfnamefont {S.}~\bibnamefont {Plumari}}, \
  and\ \bibinfo {author} {\bibfnamefont {V.}~\bibnamefont {Greco}},\ }\href
  {http://arxiv.org/abs/1407.2559} {\bibfield  {journal} {\bibinfo  {journal}
  {arXiv:1407.2559}\ ,\ \bibinfo {pages} {5}} (\bibinfo {year}
  {2014}{\natexlab{a}})},\ \Eprint {http://arxiv.org/abs/1407.2559}
  {arXiv:1407.2559} \BibitemShut {NoStop}%
\bibitem [{\citenamefont {Puglisi}\ \emph
  {et~al.}(2014{\natexlab{b}})\citenamefont {Puglisi}, \citenamefont
  {Plumari},\ and\ \citenamefont {Greco}}]{Puglisi2014b}%
  \BibitemOpen
  \bibfield  {author} {\bibinfo {author} {\bibfnamefont {A.}~\bibnamefont
  {Puglisi}}, \bibinfo {author} {\bibfnamefont {S.}~\bibnamefont {Plumari}}, \
  and\ \bibinfo {author} {\bibfnamefont {V.}~\bibnamefont {Greco}},\ }\href
  {\doibase 10.1103/PhysRevD.90.114009} {\bibfield  {journal} {\bibinfo
  {journal} {Phys. Rev. D}\ }\textbf {\bibinfo {volume} {90}},\ \bibinfo
  {pages} {114009} (\bibinfo {year} {2014}{\natexlab{b}})},\ \Eprint
  {http://arxiv.org/abs/1408.7043} {arXiv:1408.7043} \BibitemShut {NoStop}%
\bibitem [{\citenamefont {Denicol}\ \emph
  {et~al.}(2012{\natexlab{a}})\citenamefont {Denicol}, \citenamefont {Niemi},
  \citenamefont {Molnar},\ and\ \citenamefont {Rischke}}]{Denicol2012b}%
  \BibitemOpen
  \bibfield  {author} {\bibinfo {author} {\bibfnamefont {G.~S.}\ \bibnamefont
  {Denicol}}, \bibinfo {author} {\bibfnamefont {H.}~\bibnamefont {Niemi}},
  \bibinfo {author} {\bibfnamefont {E.}~\bibnamefont {Molnar}}, \ and\ \bibinfo
  {author} {\bibfnamefont {D.~H.}\ \bibnamefont {Rischke}},\ }\href
  {http://arxiv.org/abs/1202.4551} {\bibfield  {journal} {\bibinfo  {journal}
  {Physical Review D}\ }\textbf {\bibinfo {volume} {85}},\ \bibinfo {pages}
  {114047} (\bibinfo {year} {2012}{\natexlab{a}})},\ \Eprint
  {http://arxiv.org/abs/1202.4551} {arXiv:1202.4551} \BibitemShut {NoStop}%
\bibitem [{\citenamefont {Denicol}\ \emph
  {et~al.}(2012{\natexlab{b}})\citenamefont {Denicol}, \citenamefont
  {Moln\'{a}r}, \citenamefont {Niemi},\ and\ \citenamefont
  {Rischke}}]{Denicol2012}%
  \BibitemOpen
  \bibfield  {author} {\bibinfo {author} {\bibfnamefont {G.~S.}\ \bibnamefont
  {Denicol}}, \bibinfo {author} {\bibfnamefont {E.}~\bibnamefont {Moln\'{a}r}},
  \bibinfo {author} {\bibfnamefont {H.}~\bibnamefont {Niemi}}, \ and\ \bibinfo
  {author} {\bibfnamefont {D.~H.}\ \bibnamefont {Rischke}},\ }\href {\doibase
  10.1140/epja/i2012-12170-x} {\bibfield  {journal} {\bibinfo  {journal} {The
  European Physical Journal A}\ }\textbf {\bibinfo {volume} {48}},\ \bibinfo
  {pages} {170} (\bibinfo {year} {2012}{\natexlab{b}})},\ \Eprint
  {http://arxiv.org/abs/1206.1554} {arXiv:1206.1554} \BibitemShut {NoStop}%
\bibitem [{\citenamefont {Denicol}\ \emph {et~al.}(2010)\citenamefont
  {Denicol}, \citenamefont {Koide},\ and\ \citenamefont
  {Rischke}}]{Denicol2010}%
  \BibitemOpen
  \bibfield  {author} {\bibinfo {author} {\bibfnamefont {G.~S.}\ \bibnamefont
  {Denicol}}, \bibinfo {author} {\bibfnamefont {T.}~\bibnamefont {Koide}}, \
  and\ \bibinfo {author} {\bibfnamefont {D.~H.}\ \bibnamefont {Rischke}},\
  }\href {\doibase 10.1103/PhysRevLett.105.162501} {\bibfield  {journal}
  {\bibinfo  {journal} {Physical Review Letters}\ }\textbf {\bibinfo {volume}
  {105}},\ \bibinfo {pages} {162501} (\bibinfo {year} {2010})},\ \Eprint
  {http://arxiv.org/abs/1004.5013} {arXiv:1004.5013} \BibitemShut {NoStop}%
\bibitem [{\citenamefont {Moore}\ and\ \citenamefont
  {Robert}(2006)}]{Moore2003}%
  \BibitemOpen
  \bibfield  {author} {\bibinfo {author} {\bibfnamefont {G.~D.}\ \bibnamefont
  {Moore}}\ and\ \bibinfo {author} {\bibfnamefont {J.~M.}\ \bibnamefont
  {Robert}},\ }\href@noop {} {\bibfield  {journal} {\bibinfo  {journal}
  {hep-ph/0607172}\ } (\bibinfo {year} {2006})},\ \Eprint
  {http://arxiv.org/abs/0607172v1} {arXiv:0607172v1} \BibitemShut {NoStop}%
\bibitem [{\citenamefont {Baym}\ and\ \citenamefont
  {Heiselberg}(1997)}]{Baym1997}%
  \BibitemOpen
  \bibfield  {author} {\bibinfo {author} {\bibfnamefont {G.}~\bibnamefont
  {Baym}}\ and\ \bibinfo {author} {\bibfnamefont {H.}~\bibnamefont
  {Heiselberg}},\ }\href {\doibase 10.1103/PhysRevD.56.5254} {\bibfield
  {journal} {\bibinfo  {journal} {Physical Review D}\ }\textbf {\bibinfo
  {volume} {56}},\ \bibinfo {pages} {11} (\bibinfo {year} {1997})},\ \Eprint
  {http://arxiv.org/abs/9704214} {arXiv:9704214} \BibitemShut {NoStop}%
\bibitem [{\citenamefont {Tuchin}(2013)}]{Tuchin2013}%
  \BibitemOpen
  \bibfield  {author} {\bibinfo {author} {\bibfnamefont {K.}~\bibnamefont
  {Tuchin}},\ }\href
  {http://downloads.hindawi.com/journals/ahep/2013/490495.pdf
  http://arxiv.org/abs/1301.0099} {\ ,\ \bibinfo {pages} {67} (\bibinfo {year}
  {2013})},\ \Eprint {http://arxiv.org/abs/1301.0099} {arXiv:1301.0099}
  \BibitemShut {NoStop}%
\bibitem [{\citenamefont {Fern\'{a}ndez-Fraile}\ and\ \citenamefont
  {Gomez~Nicola}(2006)}]{Fernandez-Fraile2006}%
  \BibitemOpen
  \bibfield  {author} {\bibinfo {author} {\bibfnamefont {D.}~\bibnamefont
  {Fern\'{a}ndez-Fraile}}\ and\ \bibinfo {author} {\bibfnamefont
  {A.}~\bibnamefont {Gomez~Nicola}},\ }\href {\doibase
  10.1103/PhysRevD.73.045025} {\bibfield  {journal} {\bibinfo  {journal}
  {Physical Review D}\ }\textbf {\bibinfo {volume} {73}},\ \bibinfo {pages}
  {045025} (\bibinfo {year} {2006})}\BibitemShut {NoStop}%
\bibitem [{\citenamefont {Cassing}\ \emph {et~al.}(2013)\citenamefont
  {Cassing}, \citenamefont {Linnyk}, \citenamefont {Steinert},\ and\
  \citenamefont {Ozvenchuk}}]{Cassing2013}%
  \BibitemOpen
  \bibfield  {author} {\bibinfo {author} {\bibfnamefont {W.}~\bibnamefont
  {Cassing}}, \bibinfo {author} {\bibfnamefont {O.}~\bibnamefont {Linnyk}},
  \bibinfo {author} {\bibfnamefont {T.}~\bibnamefont {Steinert}}, \ and\
  \bibinfo {author} {\bibfnamefont {V.}~\bibnamefont {Ozvenchuk}},\ }\href
  {\doibase 10.1103/PhysRevLett.110.182301} {\bibfield  {journal} {\bibinfo
  {journal} {Physical Review Letters}\ }\textbf {\bibinfo {volume} {110}},\
  \bibinfo {pages} {182301} (\bibinfo {year} {2013})},\ \Eprint
  {http://arxiv.org/abs/1302.0906} {arXiv:1302.0906} \BibitemShut {NoStop}%
\bibitem [{\citenamefont {Steinert}\ and\ \citenamefont
  {Cassing}(2014)}]{Steinert2013}%
  \BibitemOpen
  \bibfield  {author} {\bibinfo {author} {\bibfnamefont {T.}~\bibnamefont
  {Steinert}}\ and\ \bibinfo {author} {\bibfnamefont {W.}~\bibnamefont
  {Cassing}},\ }\href {\doibase 10.1103/PhysRevC.89.035203} {\bibfield
  {journal} {\bibinfo  {journal} {Physical Review C}\ }\textbf {\bibinfo
  {volume} {89}},\ \bibinfo {pages} {035203} (\bibinfo {year} {2014})},\
  \Eprint {http://arxiv.org/abs/1312.3189} {arXiv:1312.3189} \BibitemShut
  {NoStop}%
\bibitem [{\citenamefont {Finazzo}\ and\ \citenamefont
  {Noronha}(2014)}]{Finazzo2014}%
  \BibitemOpen
  \bibfield  {author} {\bibinfo {author} {\bibfnamefont {S.~I.}\ \bibnamefont
  {Finazzo}}\ and\ \bibinfo {author} {\bibfnamefont {J.}~\bibnamefont
  {Noronha}},\ }\href {\doibase 10.1103/PhysRevD.89.106008} {\bibfield
  {journal} {\bibinfo  {journal} {Physical Review D}\ }\textbf {\bibinfo
  {volume} {89}},\ \bibinfo {pages} {106008} (\bibinfo {year} {2014})},\
  \Eprint {http://arxiv.org/abs/1311.6675} {arXiv:1311.6675} \BibitemShut
  {NoStop}%
\bibitem [{\citenamefont {Aarts}\ \emph {et~al.}(2015)\citenamefont {Aarts},
  \citenamefont {Allton}, \citenamefont {Amato}, \citenamefont {Giudice},
  \citenamefont {Hands},\ and\ \citenamefont {Skullerud}}]{Aarts:2014JHEP}%
  \BibitemOpen
  \bibfield  {author} {\bibinfo {author} {\bibfnamefont {G.}~\bibnamefont
  {Aarts}}, \bibinfo {author} {\bibfnamefont {C.}~\bibnamefont {Allton}},
  \bibinfo {author} {\bibfnamefont {A.}~\bibnamefont {Amato}}, \bibinfo
  {author} {\bibfnamefont {P.}~\bibnamefont {Giudice}}, \bibinfo {author}
  {\bibfnamefont {S.}~\bibnamefont {Hands}}, \ and\ \bibinfo {author}
  {\bibfnamefont {J.-I.}\ \bibnamefont {Skullerud}},\ }\href {\doibase
  10.1007/JHEP02(2015)186} {\bibfield  {journal} {\bibinfo  {journal} {JHEP}\
  }\textbf {\bibinfo {volume} {02}},\ \bibinfo {pages} {186} (\bibinfo {year}
  {2015})},\ \Eprint {http://arxiv.org/abs/1412.6411} {arXiv:1412.6411}
  \BibitemShut {NoStop}%
\bibitem [{\citenamefont {Aarts}\ \emph {et~al.}(2007)\citenamefont {Aarts},
  \citenamefont {Allton}, \citenamefont {Foley}, \citenamefont {Hands},\ and\
  \citenamefont {Kim}}]{Aarts2007}%
  \BibitemOpen
  \bibfield  {author} {\bibinfo {author} {\bibfnamefont {G.}~\bibnamefont
  {Aarts}}, \bibinfo {author} {\bibfnamefont {C.}~\bibnamefont {Allton}},
  \bibinfo {author} {\bibfnamefont {J.}~\bibnamefont {Foley}}, \bibinfo
  {author} {\bibfnamefont {S.}~\bibnamefont {Hands}}, \ and\ \bibinfo {author}
  {\bibfnamefont {S.}~\bibnamefont {Kim}},\ }\href@noop {} {\bibfield
  {journal} {\bibinfo  {journal} {Physical Review Letters}\ }\textbf {\bibinfo
  {volume} {99}},\ \bibinfo {pages} {022002} (\bibinfo {year}
  {2007})}\BibitemShut {NoStop}%
\bibitem [{\citenamefont {Brandt}\ \emph {et~al.}(2013)\citenamefont {Brandt},
  \citenamefont {Francis}, \citenamefont {Meyer},\ and\ \citenamefont
  {Wittig}}]{Brandt2013}%
  \BibitemOpen
  \bibfield  {author} {\bibinfo {author} {\bibfnamefont {B.~B.}\ \bibnamefont
  {Brandt}}, \bibinfo {author} {\bibfnamefont {A.}~\bibnamefont {Francis}},
  \bibinfo {author} {\bibfnamefont {H.~B.}\ \bibnamefont {Meyer}}, \ and\
  \bibinfo {author} {\bibfnamefont {H.}~\bibnamefont {Wittig}},\ }\href
  {\doibase 10.1007/JHEP03(2013)100} {\bibfield  {journal} {\bibinfo  {journal}
  {Journal of High Energy Physics}\ }\textbf {\bibinfo {volume} {2013}},\
  \bibinfo {pages} {100} (\bibinfo {year} {2013})}\BibitemShut {NoStop}%
\bibitem [{\citenamefont {Amato}\ \emph {et~al.}(2013)\citenamefont {Amato},
  \citenamefont {Aarts}, \citenamefont {Allton}, \citenamefont {Giudice},
  \citenamefont {Hands},\ and\ \citenamefont {Skullerud}}]{Amato2013a}%
  \BibitemOpen
  \bibfield  {author} {\bibinfo {author} {\bibfnamefont {A.}~\bibnamefont
  {Amato}}, \bibinfo {author} {\bibfnamefont {G.}~\bibnamefont {Aarts}},
  \bibinfo {author} {\bibfnamefont {C.}~\bibnamefont {Allton}}, \bibinfo
  {author} {\bibfnamefont {P.}~\bibnamefont {Giudice}}, \bibinfo {author}
  {\bibfnamefont {S.}~\bibnamefont {Hands}}, \ and\ \bibinfo {author}
  {\bibfnamefont {J.-I.}\ \bibnamefont {Skullerud}},\ }\href {\doibase
  10.1103/PhysRevLett.111.172001} {\bibfield  {journal} {\bibinfo  {journal}
  {Physical Review Letters}\ }\textbf {\bibinfo {volume} {111}},\ \bibinfo
  {pages} {172001} (\bibinfo {year} {2013})},\ \Eprint
  {http://arxiv.org/abs/1307.6763} {arXiv:1307.6763} \BibitemShut {NoStop}%
\bibitem [{\citenamefont {Gupta}(2004)}]{Gupta2004}%
  \BibitemOpen
  \bibfield  {author} {\bibinfo {author} {\bibfnamefont {S.}~\bibnamefont
  {Gupta}},\ }\href {\doibase 10.1016/j.physletb.2004.05.079} {\bibfield
  {journal} {\bibinfo  {journal} {Physics Letters B}\ }\textbf {\bibinfo
  {volume} {597}},\ \bibinfo {pages} {57} (\bibinfo {year} {2004})}\BibitemShut
  {NoStop}%
\bibitem [{\citenamefont {Buividovich}\ \emph {et~al.}(2010)\citenamefont
  {Buividovich}, \citenamefont {Chernodub}, \citenamefont {Kharzeev},
  \citenamefont {Kalaydzhyan}, \citenamefont {Luschevskaya},\ and\
  \citenamefont {Polikarpov}}]{Buividovich2010}%
  \BibitemOpen
  \bibfield  {author} {\bibinfo {author} {\bibfnamefont {P.~V.}\ \bibnamefont
  {Buividovich}}, \bibinfo {author} {\bibfnamefont {M.~N.}\ \bibnamefont
  {Chernodub}}, \bibinfo {author} {\bibfnamefont {D.~E.}\ \bibnamefont
  {Kharzeev}}, \bibinfo {author} {\bibfnamefont {T.}~\bibnamefont
  {Kalaydzhyan}}, \bibinfo {author} {\bibfnamefont {E.~V.}\ \bibnamefont
  {Luschevskaya}}, \ and\ \bibinfo {author} {\bibfnamefont {M.~I.}\
  \bibnamefont {Polikarpov}},\ }\href
  {http://link.aps.org/doi/10.1103/PhysRevLett.105.132001} {\bibfield
  {journal} {\bibinfo  {journal} {Physical Review Letters}\ }\textbf {\bibinfo
  {volume} {105}},\ \bibinfo {pages} {132001} (\bibinfo {year}
  {2010})}\BibitemShut {NoStop}%
\bibitem [{\citenamefont {Burnier}\ and\ \citenamefont
  {Laine}(2012)}]{Burnier2012}%
  \BibitemOpen
  \bibfield  {author} {\bibinfo {author} {\bibfnamefont {Y.}~\bibnamefont
  {Burnier}}\ and\ \bibinfo {author} {\bibfnamefont {M.}~\bibnamefont
  {Laine}},\ }\href {\doibase 10.1140/epjc/s10052-012-1902-8} {\bibfield
  {journal} {\bibinfo  {journal} {The European Physical Journal C}\ }\textbf
  {\bibinfo {volume} {72}},\ \bibinfo {pages} {1} (\bibinfo {year}
  {2012})}\BibitemShut {NoStop}%
\bibitem [{\citenamefont {Ding}\ \emph {et~al.}(2011)\citenamefont {Ding},
  \citenamefont {Francis}, \citenamefont {Kaczmarek}, \citenamefont {Karsch},
  \citenamefont {Laermann},\ and\ \citenamefont {Soeldner}}]{Ding2011}%
  \BibitemOpen
  \bibfield  {author} {\bibinfo {author} {\bibfnamefont {H.-T.}\ \bibnamefont
  {Ding}}, \bibinfo {author} {\bibfnamefont {A.}~\bibnamefont {Francis}},
  \bibinfo {author} {\bibfnamefont {O.}~\bibnamefont {Kaczmarek}}, \bibinfo
  {author} {\bibfnamefont {F.}~\bibnamefont {Karsch}}, \bibinfo {author}
  {\bibfnamefont {E.}~\bibnamefont {Laermann}}, \ and\ \bibinfo {author}
  {\bibfnamefont {W.}~\bibnamefont {Soeldner}},\ }\href {\doibase
  10.1103/PhysRevD.83.034504} {\bibfield  {journal} {\bibinfo  {journal}
  {Physical Review D}\ }\textbf {\bibinfo {volume} {83}},\ \bibinfo {pages}
  {034504} (\bibinfo {year} {2011})}\BibitemShut {NoStop}%
\bibitem [{\citenamefont {Kaczmarek}\ and\ \citenamefont
  {M{\"u}ller}(2014)}]{Kaczmarek:2013dya}%
  \BibitemOpen
  \bibfield  {author} {\bibinfo {author} {\bibfnamefont {O.}~\bibnamefont
  {Kaczmarek}}\ and\ \bibinfo {author} {\bibfnamefont {M.}~\bibnamefont
  {M{\"u}ller}},\ }\href@noop {} {\bibfield  {journal} {\bibinfo  {journal}
  {PoS LATTICE2013}\ ,\ \bibinfo {pages} {175}} (\bibinfo {year} {2014})},\
  \Eprint {http://arxiv.org/abs/1312.5609} {arXiv:1312.5609} \BibitemShut
  {NoStop}%
\bibitem [{\citenamefont {Qin}(2015)}]{Qin2013}%
  \BibitemOpen
  \bibfield  {author} {\bibinfo {author} {\bibfnamefont {S.-X.}\ \bibnamefont
  {Qin}},\ }\href {\doibase 10.1016/j.physletb.2015.02.009} {\bibfield
  {journal} {\bibinfo  {journal} {Phys. Lett.}\ }\textbf {\bibinfo {volume}
  {B742}},\ \bibinfo {pages} {358} (\bibinfo {year} {2015})},\ \Eprint
  {http://arxiv.org/abs/1307.4587} {arXiv:1307.4587} \BibitemShut {NoStop}%
\bibitem [{\citenamefont {Marty}\ \emph {et~al.}(2013)\citenamefont {Marty},
  \citenamefont {Bratkovskaya}, \citenamefont {Cassing}, \citenamefont
  {Aichelin},\ and\ \citenamefont {Berrehrah}}]{Marty:2013ita}%
  \BibitemOpen
  \bibfield  {author} {\bibinfo {author} {\bibfnamefont {R.}~\bibnamefont
  {Marty}}, \bibinfo {author} {\bibfnamefont {E.}~\bibnamefont {Bratkovskaya}},
  \bibinfo {author} {\bibfnamefont {W.}~\bibnamefont {Cassing}}, \bibinfo
  {author} {\bibfnamefont {J.}~\bibnamefont {Aichelin}}, \ and\ \bibinfo
  {author} {\bibfnamefont {H.}~\bibnamefont {Berrehrah}},\ }\href {\doibase
  10.1103/PhysRevC.88.045204} {\bibfield  {journal} {\bibinfo  {journal} {Phys.
  Rev.}\ }\textbf {\bibinfo {volume} {C88}},\ \bibinfo {pages} {045204}
  (\bibinfo {year} {2013})},\ \Eprint {http://arxiv.org/abs/1305.7180}
  {arXiv:1305.7180} \BibitemShut {NoStop}%
\bibitem [{\citenamefont {Berrehrah}\ \emph {et~al.}(2015)\citenamefont
  {Berrehrah}, \citenamefont {Cassing}, \citenamefont {Bratkovskaya},\ and\
  \citenamefont {Steinert}}]{Berrehrah:2015vhe}%
  \BibitemOpen
  \bibfield  {author} {\bibinfo {author} {\bibfnamefont {H.}~\bibnamefont
  {Berrehrah}}, \bibinfo {author} {\bibfnamefont {W.}~\bibnamefont {Cassing}},
  \bibinfo {author} {\bibfnamefont {E.}~\bibnamefont {Bratkovskaya}}, \ and\
  \bibinfo {author} {\bibfnamefont {T.}~\bibnamefont {Steinert}},\ }\href@noop
  {} {\  (\bibinfo {year} {2015})},\ \Eprint {http://arxiv.org/abs/1512.06909}
  {arXiv:1512.06909} \BibitemShut {NoStop}%
\bibitem [{\citenamefont {Denicol}\ \emph
  {et~al.}(2011{\natexlab{a}})\citenamefont {Denicol}, \citenamefont {Noronha},
  \citenamefont {Niemi},\ and\ \citenamefont {Rischke}}]{Denicol:2011fa}%
  \BibitemOpen
  \bibfield  {author} {\bibinfo {author} {\bibfnamefont {G.~S.}\ \bibnamefont
  {Denicol}}, \bibinfo {author} {\bibfnamefont {J.}~\bibnamefont {Noronha}},
  \bibinfo {author} {\bibfnamefont {H.}~\bibnamefont {Niemi}}, \ and\ \bibinfo
  {author} {\bibfnamefont {D.~H.}\ \bibnamefont {Rischke}},\ }\href {\doibase
  10.1103/PhysRevD.83.074019} {\bibfield  {journal} {\bibinfo  {journal} {Phys.
  Rev.}\ }\textbf {\bibinfo {volume} {D83}},\ \bibinfo {pages} {074019}
  (\bibinfo {year} {2011}{\natexlab{a}})},\ \Eprint
  {http://arxiv.org/abs/1102.4780} {arXiv:1102.4780} \BibitemShut {NoStop}%
\bibitem [{\citenamefont {Denicol}\ \emph
  {et~al.}(2011{\natexlab{b}})\citenamefont {Denicol}, \citenamefont {Noronha},
  \citenamefont {Niemi},\ and\ \citenamefont {Rischke}}]{Denicol:2011ef}%
  \BibitemOpen
  \bibfield  {author} {\bibinfo {author} {\bibfnamefont {G.~S.}\ \bibnamefont
  {Denicol}}, \bibinfo {author} {\bibfnamefont {J.}~\bibnamefont {Noronha}},
  \bibinfo {author} {\bibfnamefont {H.}~\bibnamefont {Niemi}}, \ and\ \bibinfo
  {author} {\bibfnamefont {D.~H.}\ \bibnamefont {Rischke}},\ }\href {\doibase
  10.1088/0954-3899/38/12/124177} {\bibfield  {journal} {\bibinfo  {journal}
  {J. Phys.}\ }\textbf {\bibinfo {volume} {G38}},\ \bibinfo {pages} {124177}
  (\bibinfo {year} {2011}{\natexlab{b}})},\ \Eprint
  {http://arxiv.org/abs/1108.6230} {arXiv:1108.6230} \BibitemShut {NoStop}%
\bibitem [{\citenamefont {Denicol}\ \emph {et~al.}(2013)\citenamefont
  {Denicol}, \citenamefont {Gale}, \citenamefont {Jeon},\ and\ \citenamefont
  {Noronha}}]{Denicol:2013nua}%
  \BibitemOpen
  \bibfield  {author} {\bibinfo {author} {\bibfnamefont {G.~S.}\ \bibnamefont
  {Denicol}}, \bibinfo {author} {\bibfnamefont {C.}~\bibnamefont {Gale}},
  \bibinfo {author} {\bibfnamefont {S.}~\bibnamefont {Jeon}}, \ and\ \bibinfo
  {author} {\bibfnamefont {J.}~\bibnamefont {Noronha}},\ }\href {\doibase
  10.1103/PhysRevC.88.064901} {\bibfield  {journal} {\bibinfo  {journal} {Phys.
  Rev. C}\ }\textbf {\bibinfo {volume} {88}},\ \bibinfo {pages} {064901}
  (\bibinfo {year} {2013})},\ \Eprint {http://arxiv.org/abs/1308.1923}
  {arXiv:1308.1923} \BibitemShut {NoStop}%
\bibitem [{\citenamefont {Groot}\ \emph {et~al.}(1980)\citenamefont {Groot},
  \citenamefont {van Leeuwen},\ and\ \citenamefont {{van Weert
  Ch.G.}}}]{DeGroot}%
  \BibitemOpen
  \bibfield  {author} {\bibinfo {author} {\bibfnamefont {S.~R.}\ \bibnamefont
  {Groot}}, \bibinfo {author} {\bibfnamefont {W.}~\bibnamefont {van Leeuwen}},
  \ and\ \bibinfo {author} {\bibnamefont {{van Weert Ch.G.}}},\ }\href@noop {}
  {\emph {\bibinfo {title} {{Relativistic Kinetic Theory-Principles and
  Applictions}}}}\ (\bibinfo  {publisher} {North-Holland, Amsterdam},\ \bibinfo
  {year} {1980})\BibitemShut {NoStop}%
\bibitem [{\citenamefont {Molnar}\ and\ \citenamefont
  {Wolff}(2014)}]{Molnar:2014fva}%
  \BibitemOpen
  \bibfield  {author} {\bibinfo {author} {\bibfnamefont {D.}~\bibnamefont
  {Molnar}}\ and\ \bibinfo {author} {\bibfnamefont {Z.}~\bibnamefont {Wolff}},\
  }\href@noop {} {\  (\bibinfo {year} {2014})},\ \Eprint
  {http://arxiv.org/abs/1404.7850} {arXiv:1404.7850} \BibitemShut {NoStop}%
\bibitem [{\citenamefont {Bazavov}\ \emph {et~al.}(2012)\citenamefont {Bazavov}
  \emph {et~al.}}]{Bazavov:2011nk}%
  \BibitemOpen
  \bibfield  {author} {\bibinfo {author} {\bibfnamefont {A.}~\bibnamefont
  {Bazavov}} \emph {et~al.},\ }\href {\doibase 10.1103/PhysRevD.85.054503}
  {\bibfield  {journal} {\bibinfo  {journal} {Phys. Rev. D}\ }\textbf {\bibinfo
  {volume} {85}},\ \bibinfo {pages} {054503} (\bibinfo {year} {2012})},\
  \Eprint {http://arxiv.org/abs/1111.1710} {arXiv:1111.1710} \BibitemShut
  {NoStop}%
\bibitem [{\citenamefont {Huot}\ \emph {et~al.}(2006)\citenamefont {Huot},
  \citenamefont {Kovtun}, \citenamefont {Moore}, \citenamefont {Starinets},\
  and\ \citenamefont {Yaffe}}]{Starinets2006}%
  \BibitemOpen
  \bibfield  {author} {\bibinfo {author} {\bibfnamefont {S.~C.}\ \bibnamefont
  {Huot}}, \bibinfo {author} {\bibfnamefont {P.}~\bibnamefont {Kovtun}},
  \bibinfo {author} {\bibfnamefont {G.~D.}\ \bibnamefont {Moore}}, \bibinfo
  {author} {\bibfnamefont {A.}~\bibnamefont {Starinets}}, \ and\ \bibinfo
  {author} {\bibfnamefont {L.~G.}\ \bibnamefont {Yaffe}},\ }\href {\doibase
  10.1088/1126-6708/2006/12/015} {\bibfield  {journal} {\bibinfo  {journal}
  {Journal of High Energy Physics}\ }\textbf {\bibinfo {volume} {2006}},\
  \bibinfo {pages} {015} (\bibinfo {year} {2006})},\ \Eprint
  {http://arxiv.org/abs/0607237} {arXiv:0607237 [hep-th]} \BibitemShut
  {NoStop}%
\bibitem [{\citenamefont {Rougemont}\ \emph {et~al.}(2015)\citenamefont
  {Rougemont}, \citenamefont {Noronha},\ and\ \citenamefont
  {Noronha-Hostler}}]{Noronha15}%
  \BibitemOpen
  \bibfield  {author} {\bibinfo {author} {\bibfnamefont {R.}~\bibnamefont
  {Rougemont}}, \bibinfo {author} {\bibfnamefont {J.}~\bibnamefont {Noronha}},
  \ and\ \bibinfo {author} {\bibfnamefont {J.}~\bibnamefont
  {Noronha-Hostler}},\ }\href {\doibase 10.1103/PhysRevLett.115.202301}
  {\bibfield  {journal} {\bibinfo  {journal} {Phys. Rev. Lett.}\ }\textbf
  {\bibinfo {volume} {115}},\ \bibinfo {pages} {202301} (\bibinfo {year}
  {2015})},\ \Eprint {http://arxiv.org/abs/1507.06972} {arXiv:1507.06972}
  \BibitemShut {NoStop}%
\bibitem [{\citenamefont {Cassing}\ \emph {et~al.}(1990)\citenamefont
  {Cassing}, \citenamefont {Metag}, \citenamefont {Mosel},\ and\ \citenamefont
  {Niita}}]{Cassing:1990dr}%
  \BibitemOpen
  \bibfield  {author} {\bibinfo {author} {\bibfnamefont {W.}~\bibnamefont
  {Cassing}}, \bibinfo {author} {\bibfnamefont {V.}~\bibnamefont {Metag}},
  \bibinfo {author} {\bibfnamefont {U.}~\bibnamefont {Mosel}}, \ and\ \bibinfo
  {author} {\bibfnamefont {K.}~\bibnamefont {Niita}},\ }\href {\doibase
  10.1016/0370-1573(90)90164-W} {\bibfield  {journal} {\bibinfo  {journal}
  {Phys. Rept.}\ }\textbf {\bibinfo {volume} {188}},\ \bibinfo {pages} {363}
  (\bibinfo {year} {1990})}\BibitemShut {NoStop}%
\bibitem [{\citenamefont {Bass}\ \emph {et~al.}(1998)\citenamefont {Bass} \emph
  {et~al.}}]{UrQMD1}%
  \BibitemOpen
  \bibfield  {author} {\bibinfo {author} {\bibfnamefont {S.~A.}\ \bibnamefont
  {Bass}} \emph {et~al.},\ }\href {\doibase 10.1016/S0146-6410(98)00058-1}
  {\bibfield  {journal} {\bibinfo  {journal} {Prog. Part. Nucl. Phys.}\
  }\textbf {\bibinfo {volume} {41}},\ \bibinfo {pages} {255} (\bibinfo {year}
  {1998})},\ \bibinfo {note} {[Prog. Part. Nucl. Phys.41,225(1998)]},\ \Eprint
  {http://arxiv.org/abs/nucl-th/9803035} {arXiv:nucl-th/9803035} \BibitemShut
  {NoStop}%
\bibitem [{\citenamefont {Bleicher}\ \emph {et~al.}(1999)\citenamefont
  {Bleicher}, \citenamefont {Zabrodin}, \citenamefont {Spieles}, \citenamefont
  {Bass}, \citenamefont {Ernst}, \citenamefont {Soff}, \citenamefont {Bravina},
  \citenamefont {Belkacem}, \citenamefont {Weber}, \citenamefont {Stöcker},\
  and\ \citenamefont {Greiner}}]{UrQMD2}%
  \BibitemOpen
  \bibfield  {author} {\bibinfo {author} {\bibfnamefont {M.}~\bibnamefont
  {Bleicher}}, \bibinfo {author} {\bibfnamefont {E.}~\bibnamefont {Zabrodin}},
  \bibinfo {author} {\bibfnamefont {C.}~\bibnamefont {Spieles}}, \bibinfo
  {author} {\bibfnamefont {S.~A.}\ \bibnamefont {Bass}}, \bibinfo {author}
  {\bibfnamefont {C.}~\bibnamefont {Ernst}}, \bibinfo {author} {\bibfnamefont
  {S.}~\bibnamefont {Soff}}, \bibinfo {author} {\bibfnamefont {L.}~\bibnamefont
  {Bravina}}, \bibinfo {author} {\bibfnamefont {M.}~\bibnamefont {Belkacem}},
  \bibinfo {author} {\bibfnamefont {H.}~\bibnamefont {Weber}}, \bibinfo
  {author} {\bibfnamefont {H.}~\bibnamefont {Stöcker}}, \ and\ \bibinfo
  {author} {\bibfnamefont {W.}~\bibnamefont {Greiner}},\ }\href
  {http://stacks.iop.org/0954-3899/25/i=9/a=308} {\bibfield  {journal}
  {\bibinfo  {journal} {Journal of Physics G: Nuclear and Particle Physics}\
  }\textbf {\bibinfo {volume} {25}},\ \bibinfo {pages} {1859} (\bibinfo {year}
  {1999})}\BibitemShut {NoStop}%
\bibitem [{\citenamefont {Buss}\ \emph {et~al.}(2012)\citenamefont {Buss},
  \citenamefont {Gaitanos}, \citenamefont {Gallmeister}, \citenamefont {van
  Hees}, \citenamefont {Kaskulov}, \citenamefont {Lalakulich}, \citenamefont
  {Larionov}, \citenamefont {Leitner}, \citenamefont {Weil},\ and\
  \citenamefont {Mosel}}]{GiBUU}%
  \BibitemOpen
  \bibfield  {author} {\bibinfo {author} {\bibfnamefont {O.}~\bibnamefont
  {Buss}}, \bibinfo {author} {\bibfnamefont {T.}~\bibnamefont {Gaitanos}},
  \bibinfo {author} {\bibfnamefont {K.}~\bibnamefont {Gallmeister}}, \bibinfo
  {author} {\bibfnamefont {H.}~\bibnamefont {van Hees}}, \bibinfo {author}
  {\bibfnamefont {M.}~\bibnamefont {Kaskulov}}, \bibinfo {author}
  {\bibfnamefont {O.}~\bibnamefont {Lalakulich}}, \bibinfo {author}
  {\bibfnamefont {A.~B.}\ \bibnamefont {Larionov}}, \bibinfo {author}
  {\bibfnamefont {T.}~\bibnamefont {Leitner}}, \bibinfo {author} {\bibfnamefont
  {J.}~\bibnamefont {Weil}}, \ and\ \bibinfo {author} {\bibfnamefont
  {U.}~\bibnamefont {Mosel}},\ }\href {\doibase 10.1016/j.physrep.2011.12.001}
  {\bibfield  {journal} {\bibinfo  {journal} {Phys. Rept.}\ }\textbf {\bibinfo
  {volume} {512}},\ \bibinfo {pages} {1} (\bibinfo {year} {2012})},\ \Eprint
  {http://arxiv.org/abs/1106.1344} {arXiv:1106.1344} \BibitemShut {NoStop}%
\bibitem [{\citenamefont {Olive}\ \emph {et~al.}(2014)\citenamefont {Olive}
  \emph {et~al.}}]{PDG}%
  \BibitemOpen
  \bibfield  {author} {\bibinfo {author} {\bibfnamefont {K.~A.}\ \bibnamefont
  {Olive}} \emph {et~al.} (\bibinfo {collaboration} {Particle Data Group}),\
  }\href {\doibase 10.1088/1674-1137/38/9/090001} {\bibfield  {journal}
  {\bibinfo  {journal} {Chin. Phys.}\ }\textbf {\bibinfo {volume} {C38}},\
  \bibinfo {pages} {090001} (\bibinfo {year} {2014})}\BibitemShut {NoStop}%
\end{thebibliography}%

\end{document}